\documentclass{elsart}

\usepackage{graphicx}

\usepackage{amssymb}

\begin{document}

\begin{frontmatter}

\title{Event Reconstruction in the PHENIX Central Arm Spectrometers}

\author[BNL]{J. T. Mitchell},
\author[KEK]{Y. Akiba},
\author[SUBA]{L. Aphecetche},
\author[SB]{R. Averbeck},
\author[ORNL]{T. C. Awes},
\author[PNPI]{V. Baublis},
\author[RIKEN]{A. Bazilevsky\thanksref{Prov2}},
\author[LANL]{M. J. Bennett},
\author[MUN]{H. Buesching},
\author[SB]{J. Burward-Hoy},
\author[SB]{S. Butsyk\thanksref{PNPI2}},
\author[Col]{M. Chiu},
\author[SB]{T. Christ},
\author[BNL]{T. Chujo\thanksref{TSU2}},
\author[ISU]{P. Constantin},
\author[BNL]{G. David},
\author[Prov]{A. Denisov},
\author[SB]{A. Drees},
\author[LANL]{A. G. Hansen},
\author[SB]{T. K. Hemmick},
\author[SB]{J. Jia},
\author[LLNL]{S. C. Johnson\thanksref{SB2}},
\author[BNL]{E. Kistenev},
\author[TSU]{A. Kiyomichi},
\author[Hiro]{T. Kohama},
\author[ISU]{J. G. Lajoie},
\author[SB]{J. Lauret},
\author[ISU]{A. Lebedev},
\author[VANDY]{C. F. Maguire},
\author[SB]{F. Messer},
\author[LUND]{P. Nilsson},
\author[BNL]{H. Ohnishi\thanksref{Hiro2}},
\author[YON]{J. Park},
\author[ISU]{M. Rosati},
\author[VANDY]{A. A. Rose},
\author[YON]{S. S. Ryu},
\author[Hiro]{A. Sakaguchi},
\author[TSU]{S. Sato\thanksref{Prom}},
\author[KEK]{K. Shigaki},
\author[LUND]{D. Silvermyr},
\author[Hiro]{T. Sugitate},
\author[LANL]{J. P. Sullivan},
\author[TSU]{M. Suzuki},
\author[LUND]{H. Tydesjo},
\author[LANL]{H. W. Van Hecke},
\author[SB]{J. Velkovska}
\author[RRC]{M. A. Volkov},
\author[BNL]{S. White}, and
\author[Weiz]{W. Xie\thanksref{UCR}}

\address[BNL]{Brookhaven National Laboratory, Upton, NY 11973-5000, USA}
\address[Col]{Columbia University, New York, NY 10027 and Nevis Laboratories, Irvington, NY 10533, USA}
\address[Prov]{Institute for High Energy Physics (IHEP), Protvino, Russia}
\address[Hiro]{Hiroshima University, Kagamiyama, Higashi-Hiroshima 739-8526, Japan}
\address[ISU]{Iowa State University, Ames, IA 50011, USA}
\address[KEK]{KEK, High Energy Accelerator Research Organization, Tsukuba-shi, Ibaraki-ken 305-0801, Japan}
\address[RRC]{Russian Research Center ``Kurchatov Institute'', Moscow, Russia}
\address[LLNL]{Lawrence Livermore National Laboratory, Livermore, CA 94550, USA}
\address[LANL]{Los Alamos National Laboratory, Los Alamos, NM 87545, USA}
\address[LUND]{Department of Physics, Lund University, Box 118, SE-221 00 Lund, Sweden}
\address[MUN]{Institut fuer Kernphysik, University of Muenster, D-48149 Muenster, Germany}
\address[ORNL]{Oak Ridge National Laboratory, Oak Ridge, TN 37831, USA}
\address[PNPI]{PNPI, Petersburg Nuclear Physics Institute, Gatchina, Russia}
\address[RIKEN]{RIKEN BNL Research Center, Brookhaven National Laboratory, Upton, NY 11973-5000 USA}
\address[SB]{State University of New York - Stony Brook, Stony Brook, NY 11794, USA}
\address[SUBA]{SUBATECH (Echole des Mines de Nantes, IN2P3/CNRS, Universite de Nantes) BP 20722-44307, Nantes-cedex 3, France}
\address[TSU]{Institute of Physics, University of Tsukuba, Tsukuba, Ibaraki 305, Japan}
\address[VANDY]{Vanderbilt University, Nashville, TN 37235, USA}
\address[Weiz]{Weizmann Institute, Rehovot 76100, Israel}
\address[YON]{Yonsei University, IPAP, Seoul 120-749, Korea}
\thanks[UCR]{Currently at University of California - Riverside, Riverside, CA 92521, USA}
\thanks[Hiro2]{Formerly at Hiroshima University, Kagamiyama, Higashi-Hiroshima 739-8526, Japan}
\thanks[RBRC]{Joint appointment at RIKEN BNL Research Center, Brookhaven National Laboratory, Upton, NY 11973-5000 USA}
\thanks[Prov2]{Formerly at Institute for High Energy Physics (IHEP), Protvino, Russia}
\thanks[SB2]{Formerly at State University of New York - Stony Brook, Stony Brook, NY 11794, USA}
\thanks[TSU2]{Currently at Institute of Physics, University of Tsukuba, Tsukuba, Ibaraki 305, Japan}
\thanks[PNPI2]{Currently at PNPI, Petersburg Nuclear Physics Institute, Gatchina, Russia}
\thanks[Prom]{Currently at BNL under the Fellowship of Research Abroad of Japan, Society for the
Promotion of Science, Tokyo 102-8471, Japan}

\date{\today}        

\begin{abstract}
The central arm spectrometers for the PHENIX experiment at the Relativistic
Heavy Ion Collider have been designed for the optimization of particle 
identification in relativistic heavy ion collisions.  The spectrometers
present a challenging environment for event reconstruction 
due to a very high track multiplicity in a complicated, focusing, 
magnetic field.  In order to meet this challenge, nine distinct detector
types are integrated for charged particle tracking, momentum reconstruction, 
and particle identification.  The techniques which have been developed for the
task of event reconstruction are described.
\end{abstract}


\end{frontmatter}

\section{Introduction}

\begin{figure}
\begin{center}
\rotatebox{-90.0}{\includegraphics*[width=11cm]{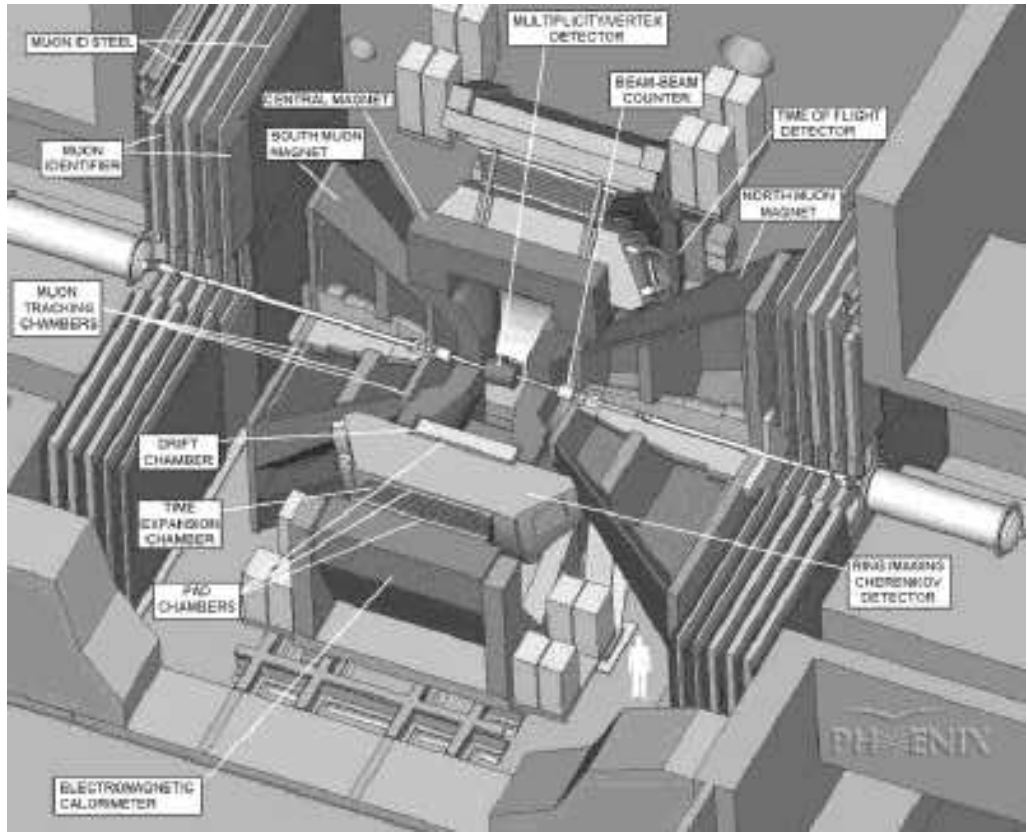}}
\end{center}
\caption{Layout of the PHENIX detector showing all four spectrometer arms within the
PHENIX experimental hall.  The collision point is in the center of the figure.}
\label{phLayout}
\end{figure}

\begin{figure}
\begin{center}
\includegraphics*[width=11cm]{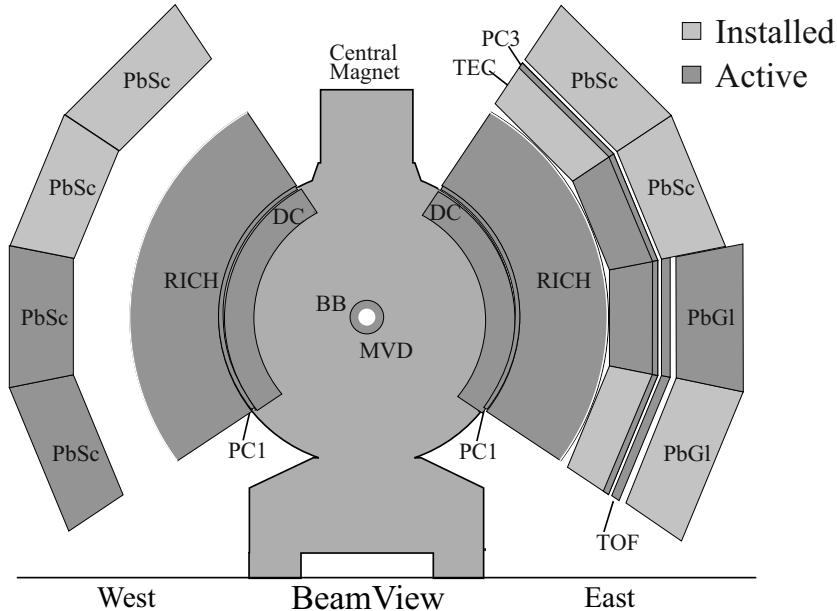}
\end{center}
\caption{Schematic diagram of the PHENIX Central Arm Spectrometers.  The
abbreviations are defined in the text.  The darker shading indicates which 
detectors were implemented for Run 2000.}
\label{phSchem}
\end{figure}

The PHENIX experiment \cite{phenix98} at the Relativistic Heavy Ion Collider 
(RHIC) is designed to measure hadrons, leptons, and photons produced in 
nucleus-nucleus, proton-nucleus, and proton-proton collisions at beam 
energies of up to 100 GeV/A with the primary goal of detecting the Quark-Gluon
Plasma (QGP) and characterizing its physical properties.  PHENIX consists of 
four spectrometers, or {\em arms} (see Fig. ~\ref{phLayout}), including two {\em muon arms} 
designed primarily for muon identification at forward rapidities, and two 
{\em central arms} focusing on hadron, electron, and photon identification near 
mid-rapidity (see Fig. ~\ref{phSchem}).  This document describes the software used for 
pattern recognition, momentum reconstruction, and particle identification for 
the two central arm spectrometers of the PHENIX detector, which began taking 
data in May 2000 until September 2000, a period which will be referred to here 
as {\em Run 2000}.

This document is organized as follows.  Section 2 provides a general overview
of the detectors in the central arm spectrometers.  Sections 3 through 10 discuss
the software for each individual detector component.  Section 11 explains the
momentum reconstruction and track model definition method.  Section 12 explains the
inter-detector association algorithm.  Section 13 explains the technique for particle
identification using the time-of-flight detector.  Section 14 summarizes what has
been discussed.

\section{The PHENIX Central Arm Spectrometers}

In order to provide a spectrometer with optimized particle identification
capabilities for various particle species, the central arm spectrometers,
which each cover $\mathnormal{\pm}$0.35 in pseudorapidity 
($\mathnormal{\eta}$) and span $90^0$ in azimuthal angle 
($\mathnormal{\phi}$), necessarily contain many components.  When there 
are differences in the two spectrometers, they will be referred to as the 
{\em east} and {\em west} arms.  The positions of the detectors will be
described in a coordinate system with the origin at the center of the
central arm magnet with the z-axis along the beam.  This section will briefly 
describe each component moving outwards radially from the beam pipe.

The innermost detector in each spectrometer is the drift chamber (DC) \cite{ria98}.
The drift chamber provides the primary vector and momentum measurement for charged
particles traversing the spectrometers.  The drift chamber consists of 40 
planes of wires arranged in 80 drift cells placed cylindrically symmetric 
about the beam line.  Each drift chamber spans $90^0$ in
$\mathnormal{\phi}$, has a radial sensitive region from 2.02 to 2.46 
meters, and covers -80 $<$ z $<$ +80 cm.  The wire planes are placed in an X-U-V 
configuration in the following order (moving outward radially): 12 X planes (X1), 
4 U planes (U1), 4 V planes (V1), 12 X planes (X2), 4 U planes (U2), and 4 V planes (V2).
The U and V planes are tilted by a small $\mathnormal{\pm}5^0$ stereo angle to allow 
for full three-dimensional track reconstruction.  The field wire design is such that the 
electron drift to each sense wire is only from one side, thus removing most 
left-right ambiguities everywhere except within 2 mm of the sense wire. The wires are 
divided electrically in the middle at $z=0$.  The occupancy for a central RHIC Au+Au collision 
is about two hits per wire.  The magnetic field bends particles in the x-y plane.  Generally, 
the drift chamber active volume lies outside of the axial magnetic field, however there is a 
residual field within the volume that provides less than $1^0$ of deflection in the bend 
plane. There is no bending, to first order, in the non-bend plane, except for tracks very close to 
the magnet pole edges or at low momentum \cite{chi96}.  The deflection of the tracks is 
sufficiently low that the track model assumption can be straight line trajectories in both 
planes.

Attached to the back of the drift chamber, at an inner inscribed radius of 
248 cm, is a single layer of pad chamber called PC1.  This is a pixel-based 
detector \cite{nil99,car97} with effective readout sizes of 8.45 mm in $z$ by 8.40 mm in 
the $x-y$ plane, providing three-dimensional space-point information for charged 
particles in the spectrometer.

Behind PC1 is a cylindrically shaped Ring Imaging Cherenkov Detector (RICH) 
\cite{aki99,aki00}.  The RICH is the primary detector for electron identification in PHENIX. It is 
a threshold Cherenkov detector with a high angular segmentation to cope with the high 
particle density expected in the most violent collisions at RHIC. There are two identical 
RICH detectors, one in each PHENIX arm. Each RICH detector consists of a gas vessel, thin 
reflection mirrors, and photon detectors made up of arrays of photo-multiplier tubes (PMT's). 
During Run 2000, {\em $CO_2$} was used as the Cherenkov radiator. Pions below 4.9 GeV/c do 
not produce a signal in the RICH, while electrons above 18 MeV/c emit Cherenkov light in the 
gas radiator. The Cherenkov photons are reflected by the mirrors in the RICH, and are 
focused onto the photon detectors. A charged particle track is identified as an electron when 
a sufficient number of RICH PMT hits are associated with it.

On the west arm only, a second pixel pad chamber, named PC2, is located
behind the RICH at an inner inscribed radius of 419 cm. This detector was not
installed for Run 2000.  PC2 has an effective readout size of 14.25 mm in $z$ 
by 13.553 mm in the $x-y$ plane.

On the east arm only, a Time Expansion Chamber (TEC) \cite{ros99} is placed behind
the RICH with a radial active area from 4.1 to 5.0 meters. One of the four installed 
TEC detector planes were instrumented for Run 2000 providing
azimuthal coverage of $\mathnormal{\phi}$ = $45^0$.  The TEC 
contains four planes of wire readout (expandable to six planes in the future).  The wires
are divided electrically at their center at z=0.  The readout electronics consist 
of a 5-bit non-linear flash ADC.  The  TEC provides 2-dimensional 
tracking in the x-y plane, contributes to electron-pion discrimination via dE/dx 
measurements, and provides high momentum resolution for high $p_t$ particles.  The 
TEC is located outside of the axial magnetic field, so a straight line track model 
assumption is valid for TEC track reconstruction.

Placed at an inner inscribed radius of 490 cm on each arm is a third layer of
pixel pad chamber, called PC3, with an effective readout size of 16.7 mm 
in $z$ by 16.0 mm in the $x-y$ plane.  Only the east arm of PC3 was installed for
Run 2000.

\begin{figure}
\begin{center}
\includegraphics*[width=6.5cm]{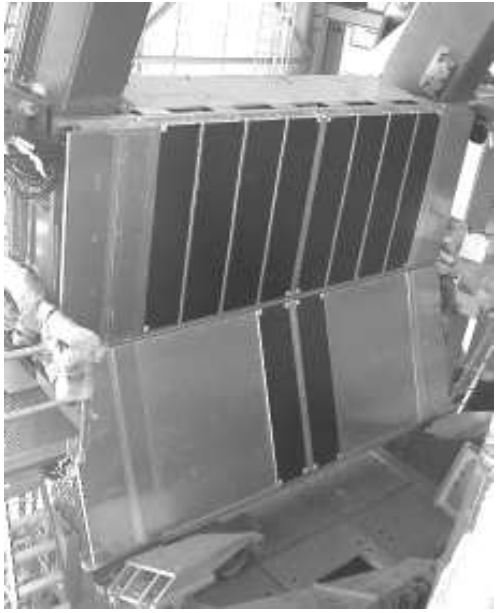}
\end{center}
\caption{The TOF detector system mounted on the PHENIX East arm in the PHENIX experimental
hall showing 10 panels of the detector.}
\label{tofpic}
\end{figure}

Located on the east arm only is the time-of-flight detector (TOF) placed at an inner 
inscribed radius of 510 cm (between PC3 and the calorimeter) and spanning $30^0$ in 
$\mathnormal{\phi}$. The TOF detector, shown mounted on the east arm in Fig. ~\ref{tofpic},
serves as the primary particle identification device for charged hadrons in PHENIX.
The TOF wall is finely segmented in order to cope with the expected highest multiplicity 
events in central Au + Au collisions, which is about 1000 charged particles per unit 
rapidity. One TOF panel contains 96 plastic scintillation counters (Bicron BC404) with 
photomultiplier tubes (Hamamatsu R3478S) at both ends of the slat. A total of 10 TOF panels, 
960 slats, and 1920 channels of PMT's were installed for Run 2000. 
Each slat is oriented along the $r$-$\mathnormal{\phi}$ direction and provides 
timing as well as longitudinal position information of particles that hit the slat.

On the east arm only, placed behind the TOF detector, are two planar arrays of
Lead-Glass calorimeters (PbGl) imported from the CERN SPS experiment WA98
\cite{pei96}, as were the TOF detectors.  Behind PC3 on the remainder of the east arm lie 
two planar arrays of Lead-Scintillator calorimeters (PbSc) \cite{dav97}.  Both of these 
arrays span $45^0$ in $\mathnormal{\phi}$ on the east arm.  Covering the full $90^0$ in 
$\mathnormal{\phi}$ on the west arm, and placed behind PC3, are four PbSc 
arrays.  The calorimeters, collectively referred to as EMC, provide photon 
identification information, particle energy, and time-of-flight measurements.

Trigger, timing, and collision centrality information is provided by a set of
detectors including a Beam-Beam Counter (BBC) \cite{ike98}, a pair of Zero-Degree 
Calorimeters (ZDC) \cite{whi98,adl01}, and a Multiplicity Vertex Detector (MVD)
\cite{ben99}.  The BBC consists of two identical sets of counters installed on 
both sides of the interaction point along the beam. Each counter is
composed of 64 PMT's equipped with a quartz Cherenkov radiator and covers the 
3.0 $< \eta <$ 3.9 region with full azimuthal coverage.  The ZDC's 
are small transverse area hadron calorimeters which measure the energy of unbound 
neutrons in small forward cones ($\mathnormal{\theta}$ $<$ 2 mr) around each beam 
axis. One ZDC is located on either side of the interaction region and consists of 3 
modules, each with a depth of 2 hadronic interaction lengths and read out by a single PMT.
Both time and amplitude are digitized for each of the 3 PMT's as well as an 
analog sum of the PMT's for each ZDC.

The MVD is a highly segmented silicon strip and pad detector. It is designed to 
measure and trigger on the number of charged particles, as well as to measure 
the collision vertex with high accuracy. The detector has a large pseudorapidity
coverage (approximately -2.64 $< \eta <$ 2.64) and full azimuthal coverage,
making it possible to study charged particle production as a function of $\eta$ and $\phi$
on an event-by-event basis.  The detector is composed of two parts: two concentric barrels,
or shells, spanning -32 $< z <$ 32 cm, and two endcaps at z=$\mathnormal{\pm}$35 cm. Each
barrel shell consists of six rows of detectors, where each row is formed by 12 panels 
of silicon strip detectors with a 200 $\mu m$ pitch. The disk-shaped endcap sections are 
each composed of six silicon pad wafers each segmented into 252 pads. When fully 
instrumented, about 35,000 channels will be read out from the MVD.

\section{Beam-Beam Counters and Zero Degree Calorimeters}

The Beam-Beam Counters (BBC) and Zero Degree Calorimeters (ZDC) are designed 
to define the start timing of the nuclear interaction for time-of-flight 
measurements, to provide the trigger signal of the nuclear interaction, and 
to provide a coarse measurement of the collision vertex along the beam axis.  The following equations 
are used to calculate the start timing ($T_0$) and the vertex position along 
the beam axis ($Z_{vertex}$) for both detectors: $T_0 = (T_1+T_2)/2$ and
$Z_{vertex} = (T_1-T_2) / 2c$, where $T_1$ and $T_2$ are the average hit timing 
for each set of BBC counters, and $c$ is the velocity of light.

The average hit timing in each BBC set is defined by taking a truncated 
average of the timing over the PMT's that have a hit (defined as a PMT with 
a valid timing signal) out of the 64 possible PMT's.  In the case that there 
is no peak structure in the hit timing distribution in either BBC set due to 
low particle multiplicity, the arrival times of the first leading particle 
from the beam collisions are used as the hit timing value.

The ZDC pulse height is converted to hadronic energy equivalent using
peripheral interactions with low neutron multiplicity in the ZDC where
single- and double-neutron peaks are clearly seen. The fitted single-neutron 
peak position is equivalent to 65 AGeV per beam for Run 2000. This fit procedure 
was performed using an {\em analog sum} channel since pedestal offset corrections 
are reduced in this channel. The offset arises because the analog-to-digital converters used 
for the BBC and ZDC are self-gating, so a minimum signal must be present to 
start the digitization.

\begin{figure}
\begin{center}
\scalebox{1.0}{\includegraphics{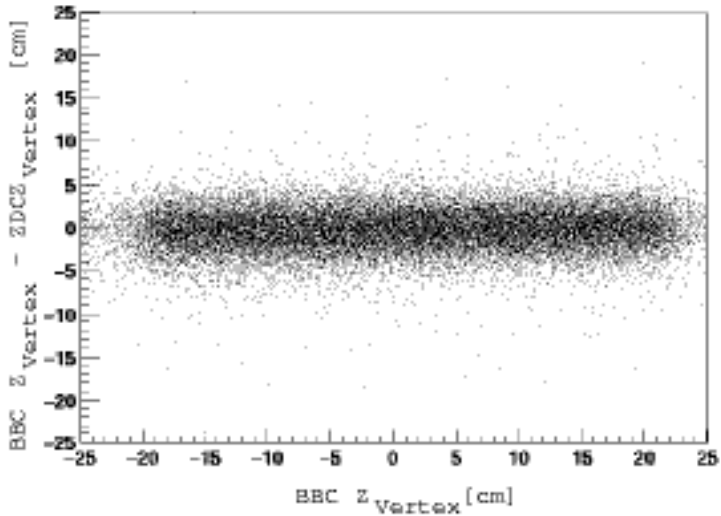}}
\end{center}
\caption{The difference in the measured BBC and ZDC $Z_{vertex}$
positions as a function of the measured BBC $Z_{vertex}$ position
from the Run 2000 data.}
\label{zvertex}
\end{figure}

The time-of-flight to the ZDC was calculated using the PMT digitized times
after correcting for pulse height slewing via a look-up table. The typical
signal rise-time is 2 nsec. The time digitizer least-count scale was
calibrated with standard delay cables. The resulting time measurements are
used to calculate $T_0$ and $Z_{vertex}$ as outlined above using a C++ 
class whose data members are stored for each event for both the BBC and ZDC.
Fig. ~\ref{zvertex} plots the difference between the measured ZDC and BBC vertex 
positions as a function of the BBC vertex position. The resolution in the
difference is 2.0 cm, which is dominated by the contribution from the 
ZDC time resolution.

\section{Multiplicity Vertex Detector}

The Multiplicity Vertex Detector (MVD) is designed to measure and trigger
on the number of charged particles produced in high multiplicity events
in Au+Au collisions and to measure the collision vertex with high accuracy.
Vertex finding is performed by the MVD barrel section where the hit information 
from the two shells of detectors can be combined and projected onto the z-axis.
Multiplicity measurements are performed by both the barrel and pad sections.
For a vertex at the center of the MVD ($z=0$), the inner shell covers the range 
-2.55 $< \eta < $ 2.55 and the Si pad detectors cover the range 
1.79 $< | \eta |  < $ 2.64 at each end of the detector. Both the inner shell and 
the pads have complete azimuthal coverage.

The MVD event processing, written in C++, is handled by an MVD event 
reconstruction object, which is created and evaluated for each event 
by the main MVD reconstruction module. The event reconstruction object 
contains data members holding the hit information for each event and the final
vertex and $dN/d\eta$ results. The content of these data members are
evaluated by a call to the internal event processing method. Strips,
pads, hit clusters, and the vertex are treated as individual objects
for the given event. The content of the cluster objects are generated
from the strip objects. Strip, pad, and cluster objects are accessed by
a vertex object class which holds the methods containing the main
reconstruction algorithms.

\begin{figure}
\begin{center}
\scalebox{1.0}{\includegraphics{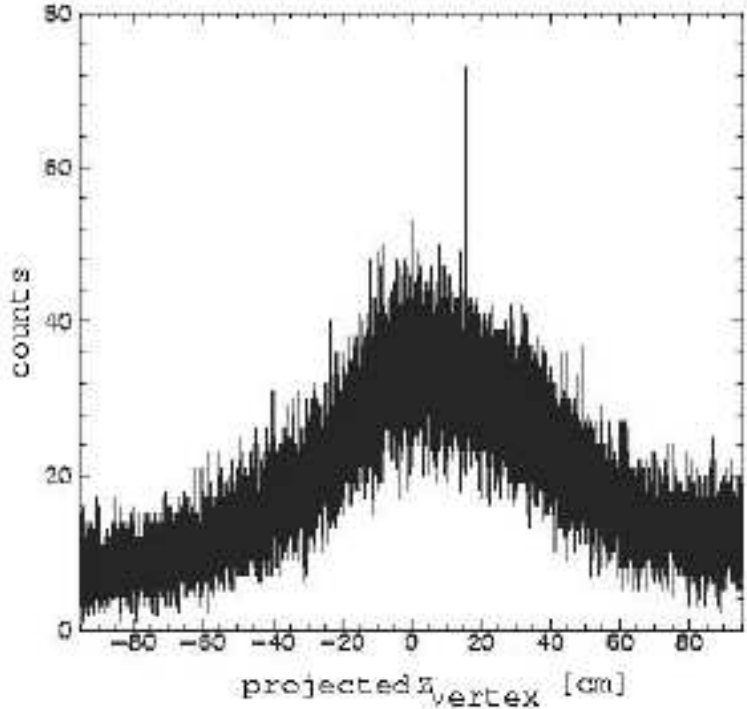}}
\end{center}
\caption{MVD vertex finding histogram for the pseudo-tracking method using
simulated central HIJING Au+Au events. The narrow peak defines the vertex.}
\label{mvdpseudo}
\end{figure}

Two different algorithms are used to find the vertex position.  The
first algorithm is denoted the {\em pseudo-tracking} method.  This vertex
algorithm takes each pair of clusters (where a cluster is a group of adjacent
channels in the same row, shell, and panel of the MVD which all have
signals above a threshold) in the inner and outer shells and treats
every combination in the same row of the MVD barrel as a potential
track. This combination of track candidates is projected back to the beam axis to
determine the vertex expected for a track which hit these two locations. For 
each of these possible vertex locations, a channel in a histogram is 
incremented. Most pairs of hits do not correspond to a real track. However, 
when all pairs are considered, the true vertex location appears more frequently 
than the other locations. A clear peak appears in the histogram corresponding 
to the vertex position, as shown in Fig. ~\ref{mvdpseudo} from simulated central Au+Au events
using the HIJING event generator \cite{wan91}.  When this algorithm finds the correct 
vertex location, it is correct within a standard deviation, $\sigma$, of about 190 $\mu$m for
simulated central events and 224 $\mu$m for simulated minimum bias events. At high occupancy 
(around 45\% in the inner shell) this method becomes more inefficient (less than 98\%) and a 
different approach must be used.

\begin{figure}
\begin{center}
\scalebox{1.0}{\includegraphics{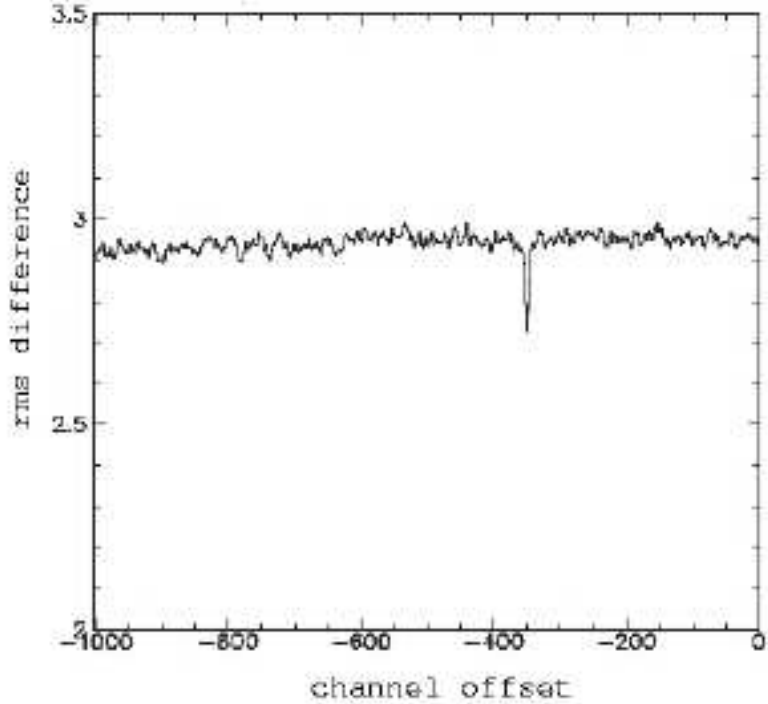}}
\end{center}
\caption{MVD vertex finding histogram for the correlation method using
simulated central HIJING Au+Au events. The narrow dip defines the vertex.}
\label{mvdcorr}
\end{figure}

The alternative vertex algorithm, used for high multiplicity events, is
the {\em correlation method}. Unlike the pseudo-tracking method, the
correlation method makes use of the ADC information from the individual channels. 
The algorithm is based upon the assumption that the pattern of energy loss per channel 
in the inner and outer shells of the MVD are similar. The pattern of ADC values in the outer 
shell of the MVD is scaled down by the ratio of the inner-to-outer shell radius and stored
in an array. The scaled pattern yields an offset proportional to the vertex location when
compared to the unscaled inner shell pattern. By finding the value of the offset which makes 
the two patterns most similar, the vertex position is found. A histogram which plots 
the difference between the pattern of hits in the inner and outer shells will show a 
sharp minimum which corresponds to the location of the vertex, as shown in Fig. ~\ref{mvdcorr} 
from central HIJING Au+Au events. This method gives the correct vertex with a standard
deviation of 233 $\mu$m for simulated central events and 250 $\mu$m for simulated minimum bias events
with an efficiency greater than 99\% for inner shell occupancies above 5\%.

The algorithms for calculating $dN/d\eta$ are different in the MVD barrel section and 
endcap sections. In the MVD barrel, only the inner shell is used. In central Au+Au 
collisions, the occupancy in the inner shell is high \cite{ben99}, 
therefore the algorithm makes no attempt to find individual hits. Instead, 
the barrel is divided into groups of 64 adjacent strips which are in the same row (azimuthal
segment). For each of these groups of channels, a $dN/d\eta$ value is calculated as 
follows: the total ADC value for all channels in this group is added up, the range of 
pseudorapidity ($\Delta\eta$) is calculated, the pseudorapidity at the center of 
the group of strips is calculated, a geometric correction (for non-normal incidence) is 
applied, the number of particles is estimated by dividing the corrected ADC signal by the 
ADC signal expected for a minimum ionizing particle ({\em mip}), and finally $dN/d\eta$ is
estimated as the number of particles divided by $\Delta\eta$.  The average $dN/d\eta$ for a set 
of events is calculated by taking the average in each bin of $\eta$.

In order to calculate $dN/d\eta$ from the MVD pads, the algorithm assumes that a 
single particle hits only one pad. The occupancy of the pad detectors is around 15-20\% 
for central events. The algorithm estimates the number of particles associated with the 
observed ADC value in an individual pad using an ADC distribution which is taken from 
the distribution corresponding to the ADC response expected for a single particle at 
normal incident angle in low multiplicity events, and the occupancy of the pad detector. 
The algorithm does not associate an integer number of particles with each hit. Instead a mean 
number of particles which would be associated with a given ADC channel is calculated. 
This number can then be converted into a $dN/d\eta$ value using the vertex location and 
the pad geometry.

\section{Pad Chambers}

\begin{figure}
\begin{center}
\includegraphics*[width=11cm]{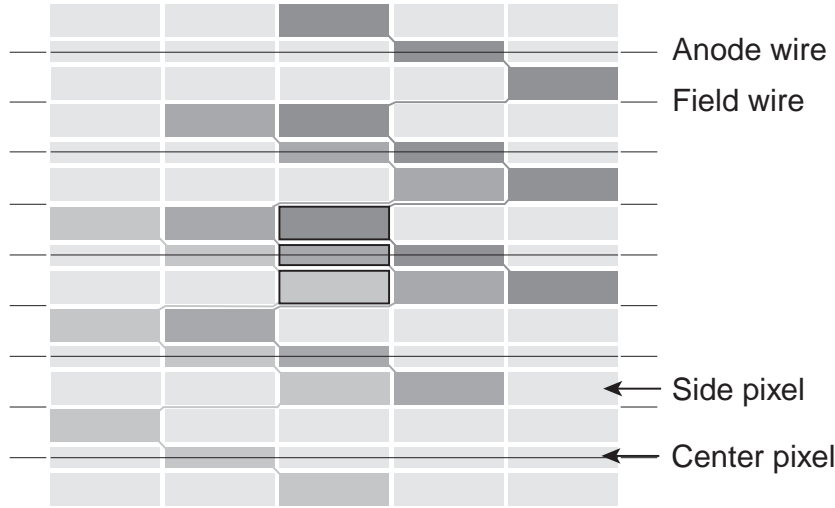}
\end{center}
\caption{The pad chamber pixel pattern.  The three pixels form a cell at the center
of the figure. Nine connected pixels form a pad.}
\label{padSchem}
\end{figure}

A new pad geometry for cathode readout of MWPC's was designed for the PHENIX pad
chambers~\cite{car97,sve99}. The basic component in this geometry is a pad consisting of
nine smaller connected copper electrodes, called pixels (Fig. ~\ref{padSchem}).
The pad is read out by a single preamplifier and discriminator. The pads are
interleaved in a repeated pattern. The area defined by three adjacent pixels belonging to
three different pads is called a cell. This arrangement saves a factor of three 
in the number of readout channels with marginal loss of performance.

Pad chamber information, which is stored in the PHENIX Objectivity database, can be
separated into three different groups: geometry, high voltage and electronics. All
groups are accessed via object-oriented classes in the offline analysis, using time 
stamps as the primary key. The geometry group holds information on the geometry parameters of 
the chambers themselves, e.g. wire-spacing, as well as survey information, while the 
high voltage group keeps track of when trips have occurred. The electronics group contains
threshold and gain calibrations, as well as information on where hot or inactive channels
are located.

\begin{figure}[htb]
\begin{minipage}[t]{140mm}
\scalebox{0.35}{\includegraphics{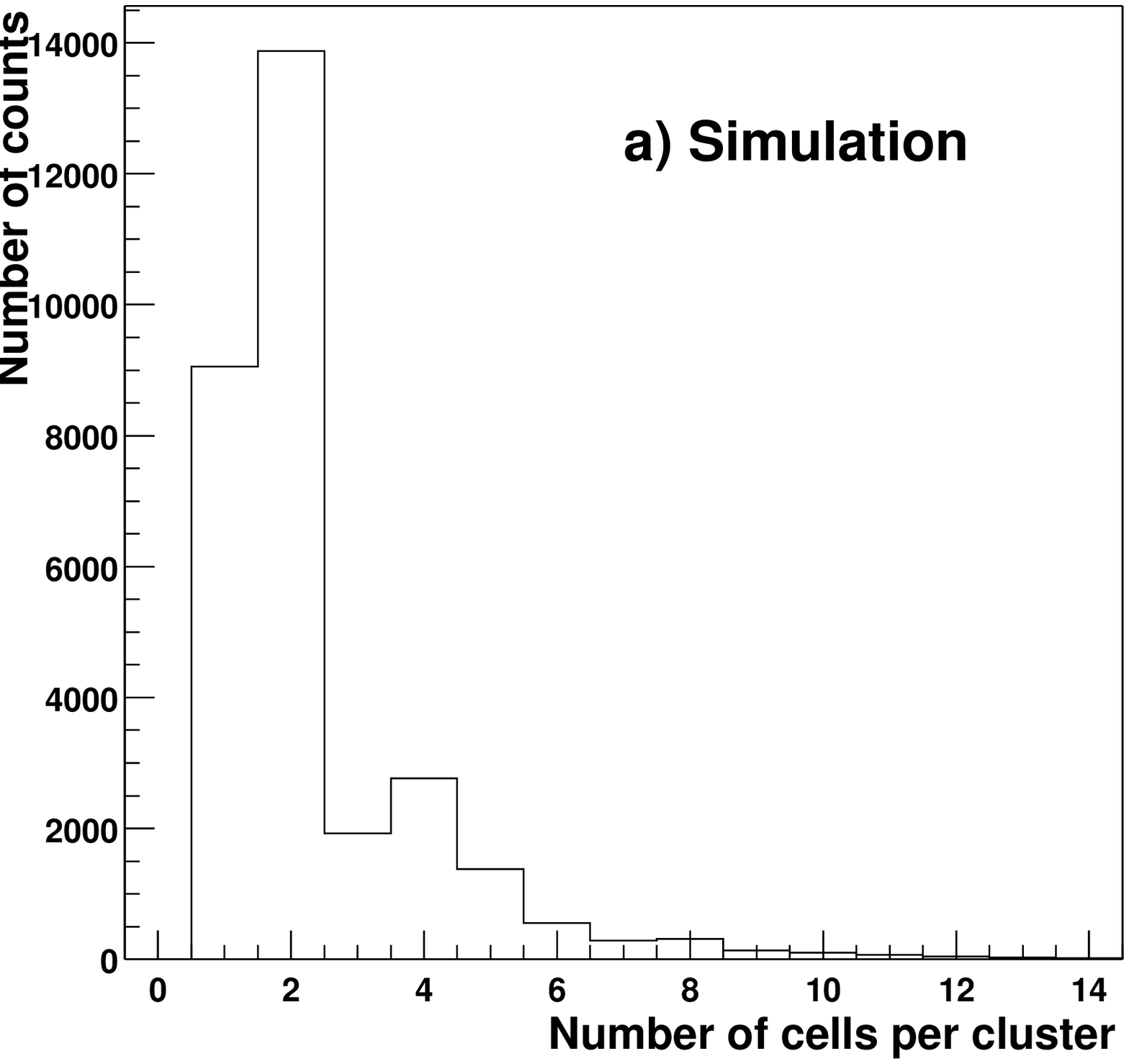}}
\hspace{\fill}
\scalebox{0.35}{\includegraphics{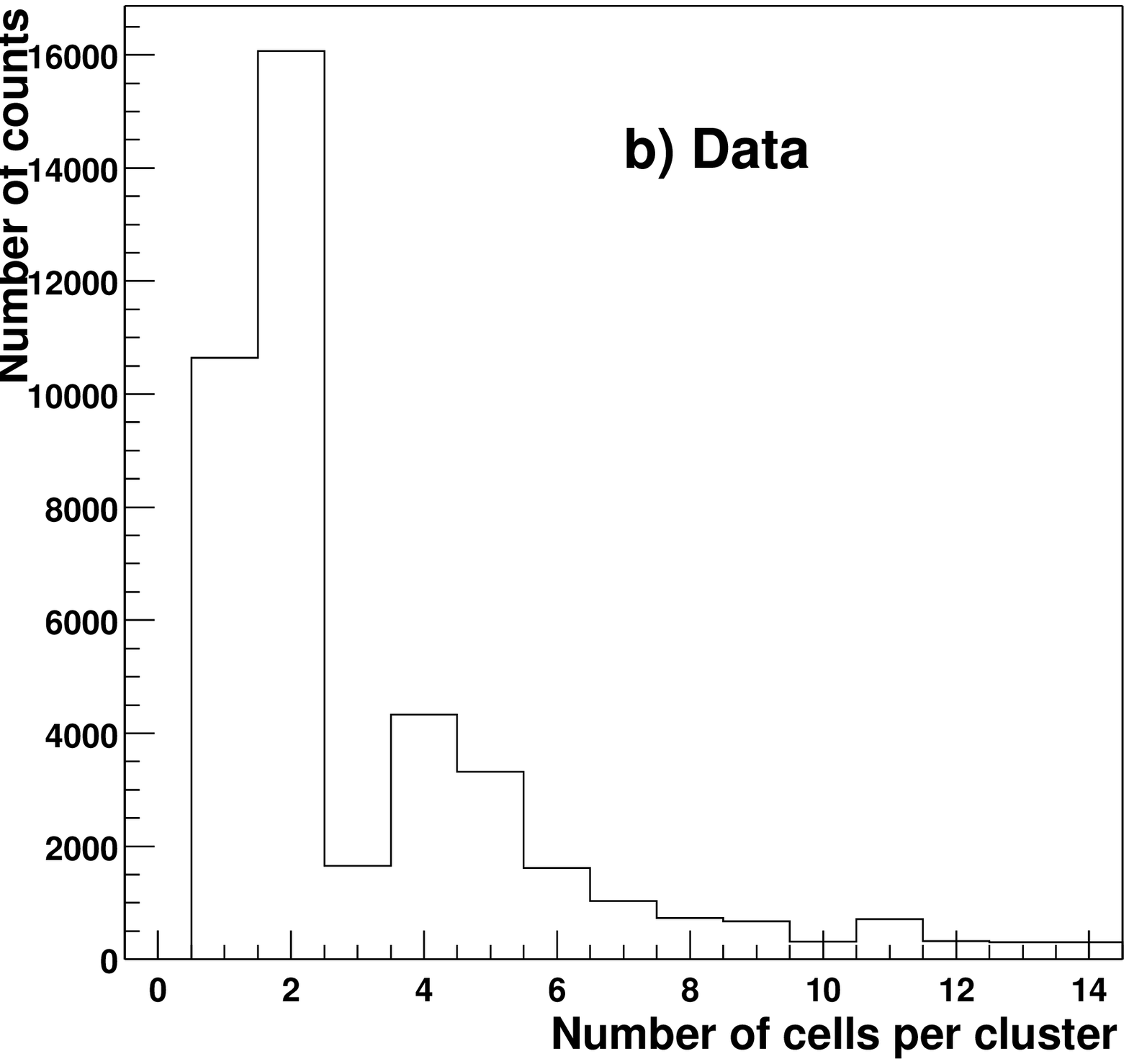}}
\vspace{0.1cm}
\caption{The distribution of the number of cells per cluster in PC3 from a) simulation
and b) data.}
\vspace{0.5cm}
\label{padFired}
\end{minipage}
\end{figure}

The list of fired pads from an event is read by the hit reconstruction algorithm,
which first translates it into a list of fired cells. The limit on the number of allowed 
inactive or hot pads in a cell is an input parameter whose default value is one, i.e. 
at least two out of three pixels in a cell have to be part of alive channels. From the cell 
list, cluster objects are built for each chamber by identifying all neighboring cells.
The size of each cluster is checked against a pre-defined table to estimate the number 
of traversing particles that created the cluster. The default settings of this table 
were determined from simulations~\cite{sve99}. If the number of particles in a cluster 
is considered to be greater than one, the cluster will be recursively split depending 
on its shape until only one-particle clusters remain. Once a cluster created by a single particle
has been extracted, the average hit position is calculated by averaging over the coordinates 
of the cells in the cluster. The entire procedure is repeated until the list of fired cells 
is empty.

\begin{figure}[htb]
\begin{minipage}[t]{140mm}
\scalebox{0.35}{\includegraphics{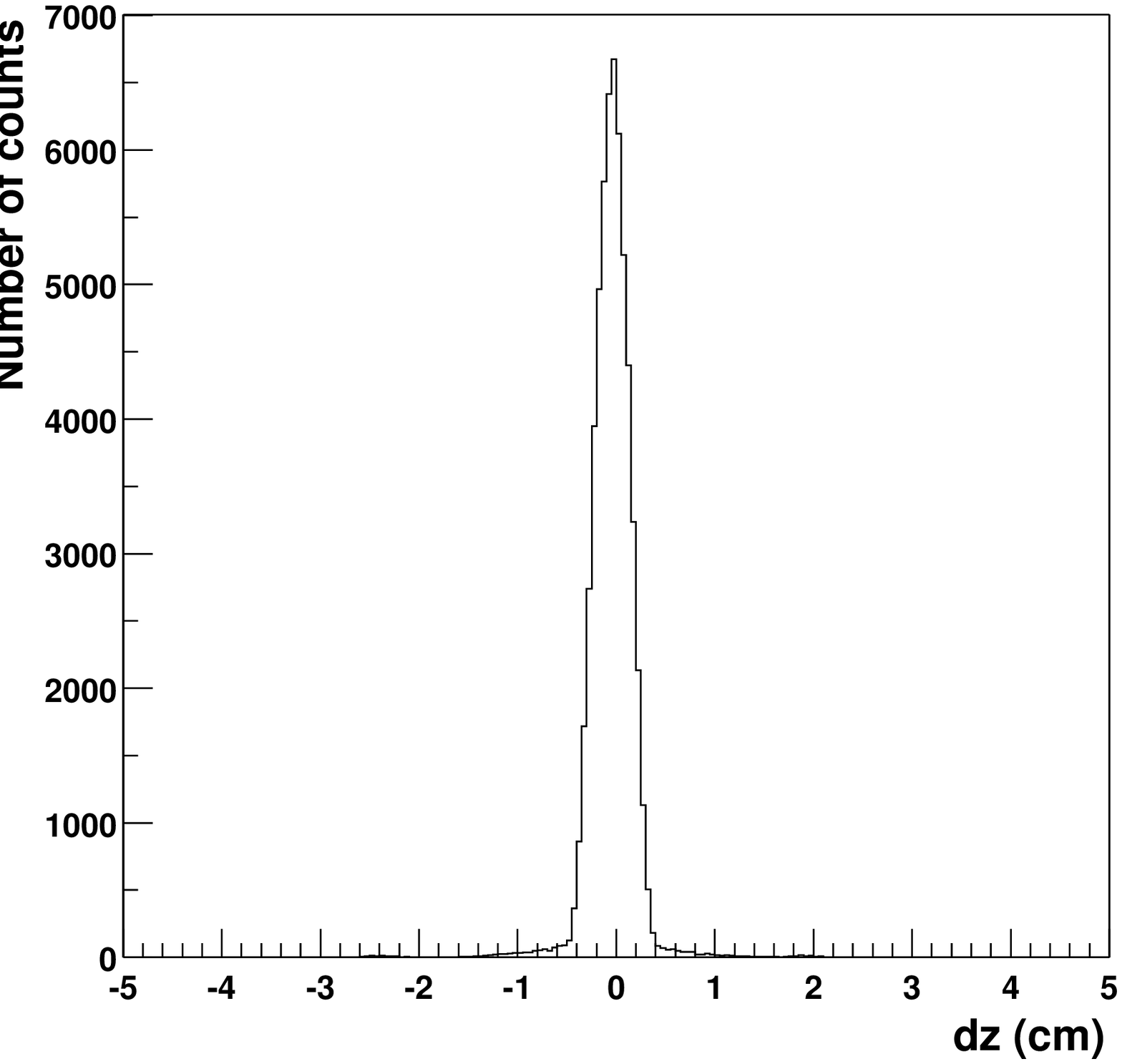}}
\hspace{\fill}
\scalebox{0.35}{\includegraphics{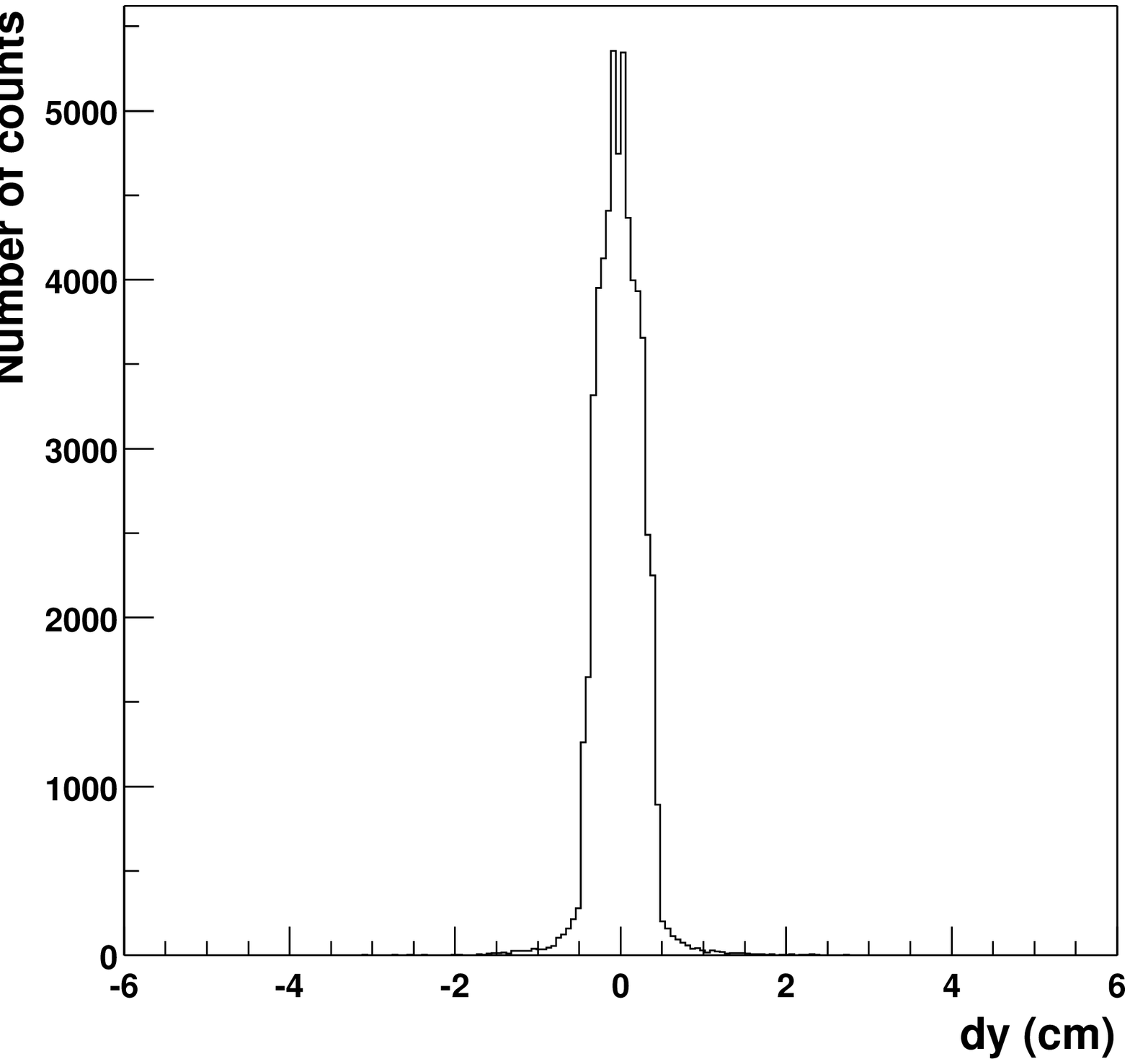}}
\vspace{0.1cm}
\caption{The simulated position resolution along the wire (left panel) and across the wire 
(right panel) in PC1.}
\vspace{0.5cm}
\label{padres}
\end{minipage}
\end{figure}

In order to determine the reconstruction efficiency, a pad chamber response simulator, 
which simulates which pads fired in an event, is utilized. Given the coordinates of a 
particle entering and exiting the pad chamber, the location of the avalanche along the
closest wire is calculated. The particle may also give rise to additional avalanches on 
neighbouring wires. The total avalanche charge is obtained from a Landau distribution. 
The induced charge distribution on the pad cathode plane is then calculated from an 
empirical formula~\cite{byu91,gat79}. Finally, when all particles of the event are processed, 
noise charge drawn from a normal distribution is added to each pad in the chamber. If the 
total collected charge on a pad exceeds the threshold value, the pad is registered as fired.
The thresholds used in the simulation were set to match the cluster sizes (Fig. ~\ref{padFired}) and 
the very high efficiency of the real chambers.

\begin{figure}[htb]
\begin{minipage}[t]{140mm}
\scalebox{0.35}{\includegraphics{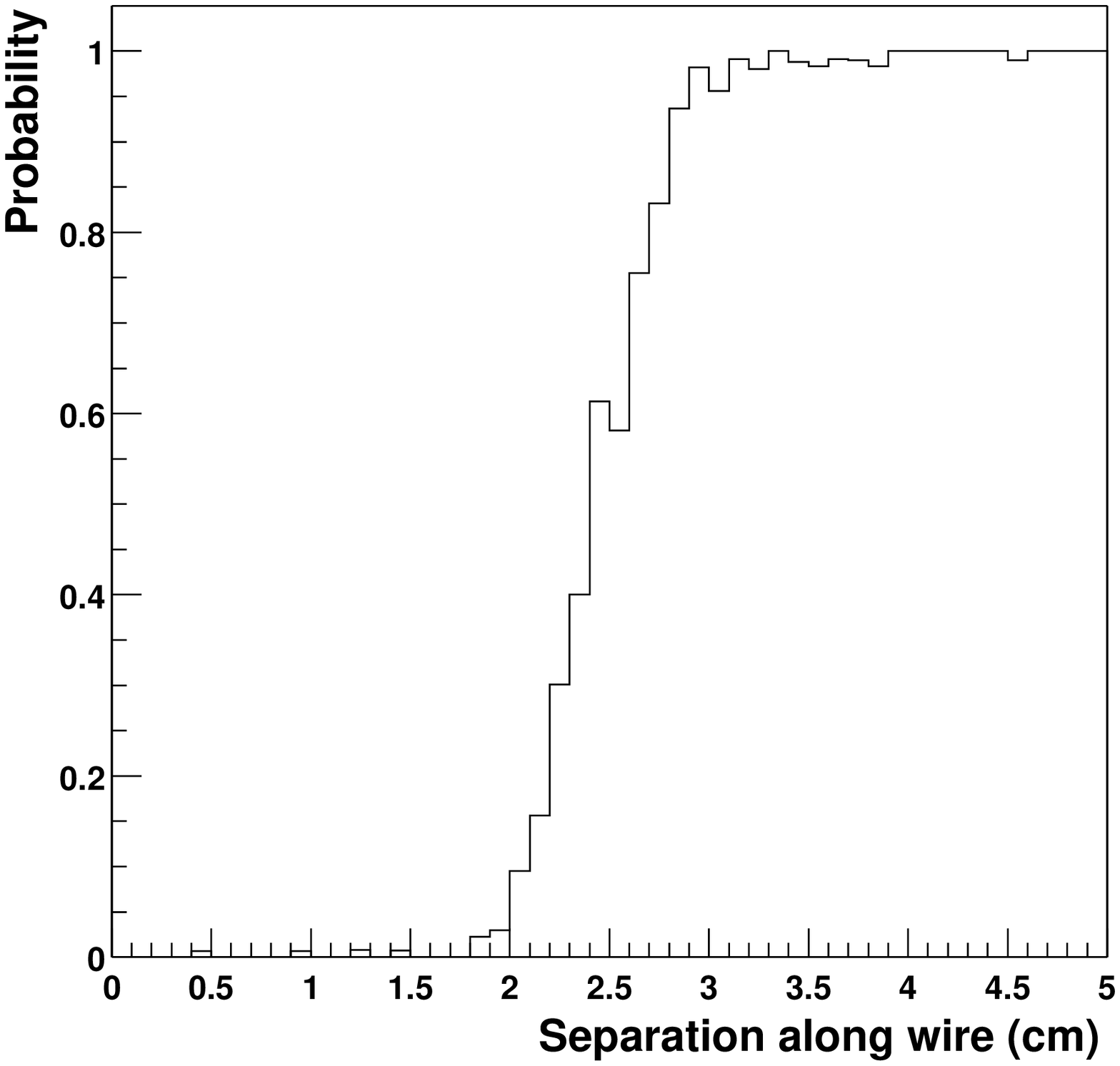}}
\hspace{\fill}
\scalebox{0.35}{\includegraphics{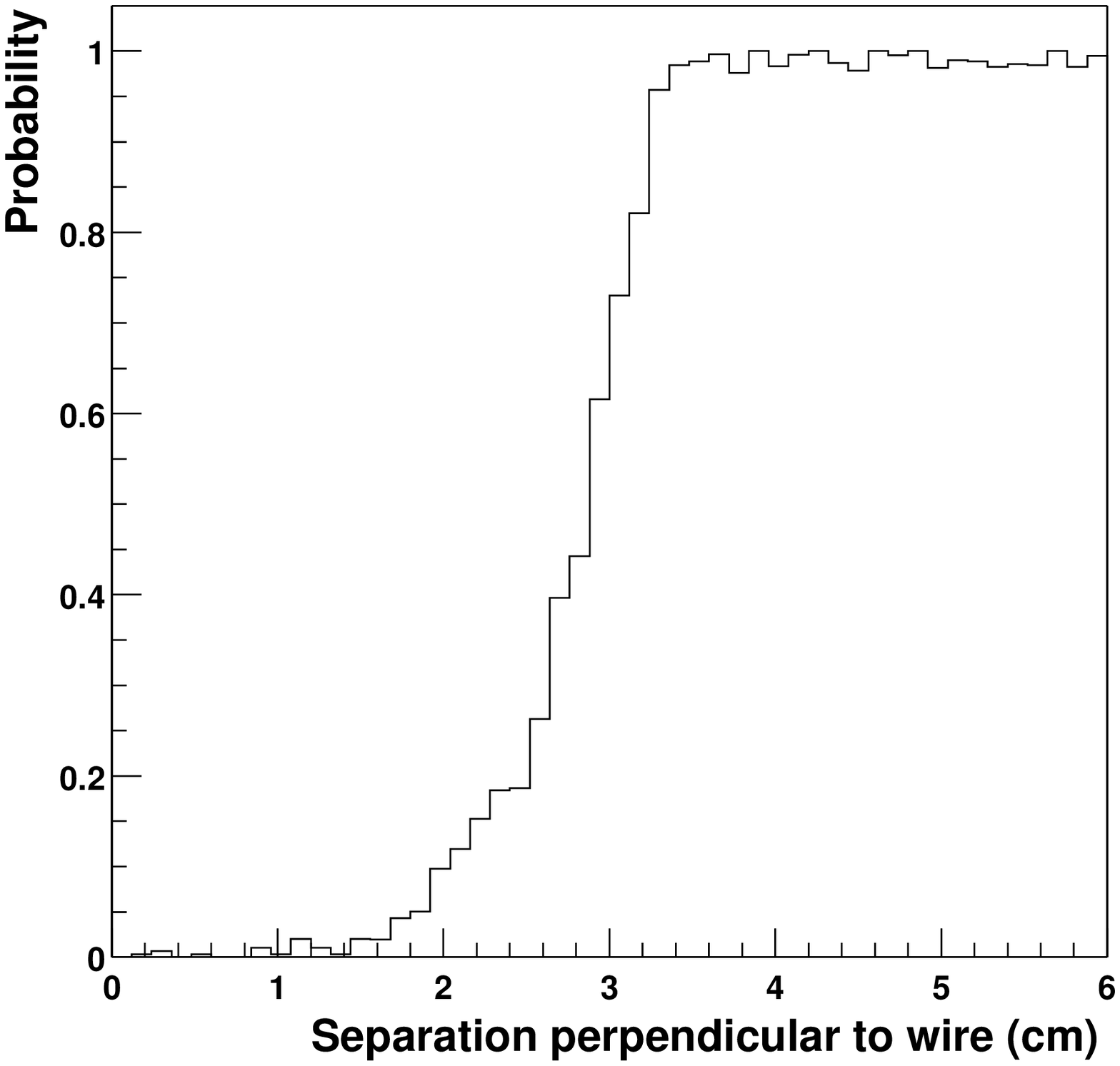}}
\vspace{0.1cm}
\caption{The probability to reconstruct two simulated particles as two separate
hits as a function of the separation of the particles in PC1 both along the wire
(left panel) and across the wire (right panel).}
\vspace{0.5cm}
\label{pad2Track}
\end{minipage}
\end{figure}

\begin{table}[h]
\begin{center}
\caption{\label{table1}Pad chamber position and two-track resolutions.}
\begin{tabular}{|l|l|l|l|} \hline
{\em Simulation Results} & {\em PC1} & {\em PC2} & {\em PC3}\\ \hline
Position resolution along wire [mm] & $1.6$ & $2.7$ & $3.2$ \\
Position resolution across wire [mm] & $2.3$ & $4.0$ & $4.8$ \\ \hline
Two-track resolution along wire [cm] & $2.4$ & $4.0$ & $5.0$ \\
Two-track resolution across wire [cm] & $2.9$ & $4.6$ & $5.3$ \\ \hline
\end{tabular}
\end{center}
\end{table}

The position resolutions, presented in Table~\ref{table1}, are determined by comparing the 
reconstructed hit positions with the positions where the original GEANT~\cite{bru94} particles 
passed through the chambers. The position resolutions for PC1 are shown in Fig. ~\ref{padres}.  
By simulating minimum ionizing particle pairs with small separation between them, the probability 
for resolving the nearby tracks was calculated. Shown in Fig. ~\ref{pad2Track} is the two-track 
resolution along and perpendicular to the wires for the PC1 chamber. The resolution is defined 
as the distance where 50\% of the tracks are separated by the reconstruction. The system can be 
optimized for two-track resolution at the expense of an increased number of ghost hits (due to 
falsely split clusters) or loss of single track efficiency by operating at lower sensitivity
(leading to reduced cluster sizes). The study presented here results in a negligible number of 
ghosts.  While the detection efficiency as measured with cosmic rays is greater than 99.5\%, 
the reconstruction efficiency is slightly lower as determined by projecting reconstructed
drift chamber tracks from minimum bias data into PC1. This results in a reconstruction efficiency 
calculated to be better than 98\%, as expected from the pad chamber simulation.

\section{Drift Chamber}

\begin{figure}
\begin{center}
\includegraphics*[width=12cm]{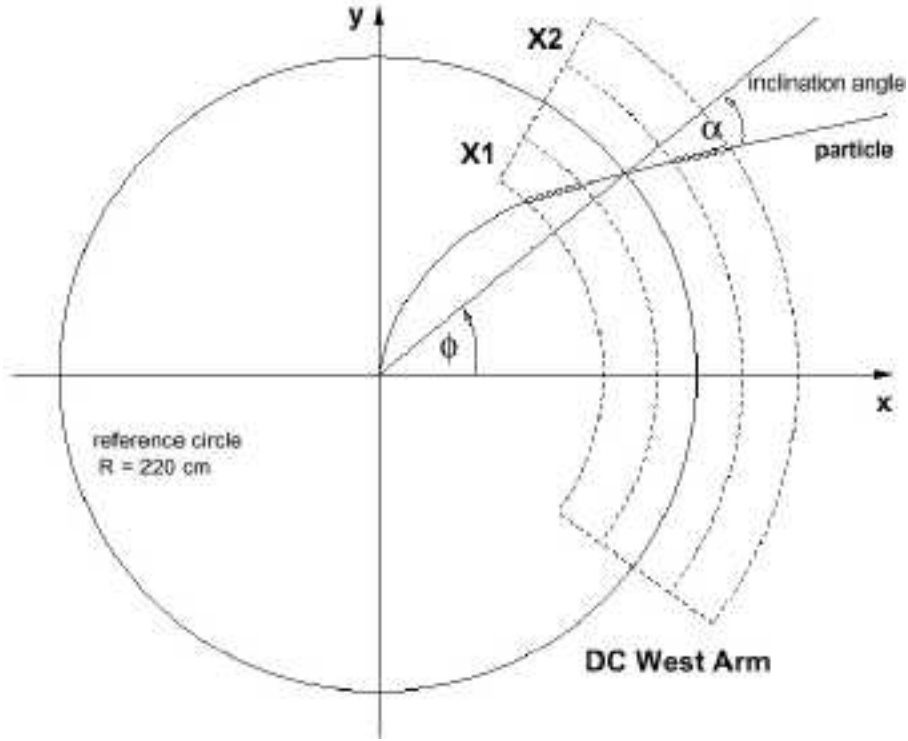}
\end{center}
\caption{Illustration of the Hough transform parameters for drift chamber track
reconstruction. The outline shows the drift chamber active volume.  The circles
represent drift chamber hits along the particle trajectory.}
\label{dchough}
\end{figure}

Track reconstruction within the drift chamber is performed using a
{\em combinatorial Hough transform} (CHT) technique \cite{ben90,ohl92}.  In this 
technique, the drift chamber hits are mapped pair-wise into a feature space defined by the 
polar angle at the intersection of the track with a reference radius near the mid-point of 
the drift chamber, $\mathnormal{\phi}$, and the inclination of the track at that point, 
$\mathnormal{\alpha}$.  The $\mathnormal{\alpha}$ variable is proportional to the inverse of the 
transverse momentum, thus facilitating limited searches for specific momentum ranges
and providing an initial guess for the momentum reconstruction procedure.
Fig. ~\ref{dchough} provides a schematic illustration of these variables.

\begin{figure}[htb]
\begin{minipage}[t]{140mm}
\begin{center}
\rotatebox{-90.0}{\scalebox{0.8}{\includegraphics{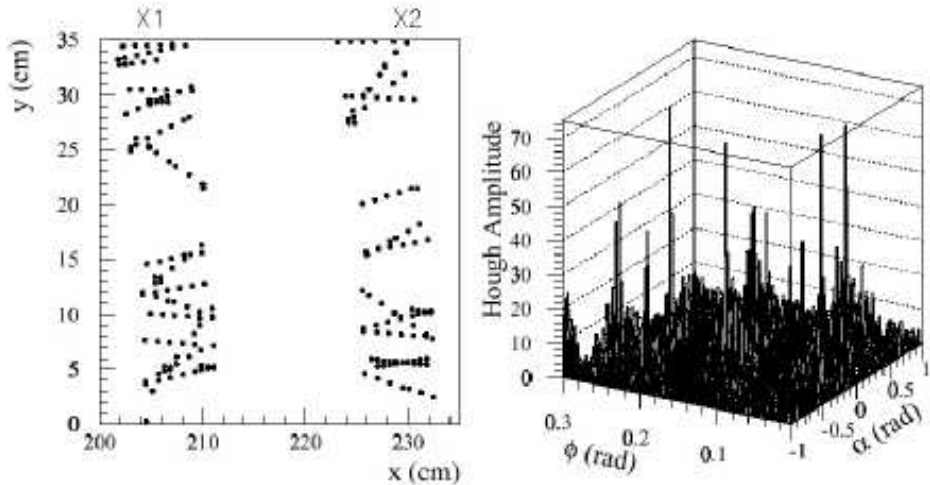}}}
\end{center}
\caption{The left panel shows simulated hits from a central Au+Au collision for a small physical
region of the drift chamber.  The right panel shows the Hough transform feature space for this
region.  Tracks appear as peaks in this plot.}
\label{dcmethod}
\end{minipage}
\end{figure}

This coordinate mapping results in a gain effect that greatly enhances the peak-to-background 
ratio in the feature space.  If {\em n} is the number of points on the track, the nominal 
peak height is $n(n-1)/2$.  In order to minimize computation time, $\mathnormal{\alpha}$ and 
$\mathnormal{\phi}$ were only calculated for hit pairs separated in azimuthal angle by physically 
reasonable amounts corresponding to $p_t$ = 150 MeV/c, comparable to the natural resolution 
of the spectrometer.  Fig. ~\ref{dcmethod} shows an example of a region of the drift chamber hits and the 
associated feature space from a simulated HIJING central RHIC Au+Au collision.  In order to
reconstruct tracks in regions where some wires have been removed for internal supports, the 
algorithm first looks for tracks that traverse both the X1 and X2 wire regions, then it looks 
for the remaining tracks that traverse the X1 or X2 regions only.

After reconstructing a track in the magnetic field bend plane, the direction of the track is
specified by $\mathnormal{\phi}$ and $\mathnormal{\alpha}$.  Track reconstruction in the 
non-bend plane is first attempted by integrating information from PC1 reconstructed clusters, 
which contain $z$ information with a resolution of 1.89 mm.  A straight line projection 
of the bend-plane track is made to the PC1 detector and a road is defined about that projection.
If there is an unambiguous PC1 association, the non-bend vector is defined by the PC1 cluster and 
the $z$-coordinate of the event vertex.  This assumption is not valid for secondary tracks, but 
those cases are later examined during momentum reconstruction.  If there is no PC1 cluster association, 
or if there are multiple PC1 association solutions, the non-bend vector is determined from 
the stereo wires.  For this procedure, a plane, called the X-plane, is defined by 
$\mathnormal{\phi}$ and $\mathnormal{\alpha}$, but extending in $z$.  The path
of the track trajectory through the X-plane is a straight line and is determined by including the 
information measured by the stereo wires.  To do this, all drift chamber hits are defined as lines 
in physical space parallel to the anode wire and perpendicular to the drift direction.  All stereo 
wire hit lines are intersected with the X-plane with the intersections belonging to 
the track defining a line in the X-plane.  Again, a Hough transform is used to extract these 
intersections by using the $z$ coordinate at which the track crosses the reference radius, 
$zed$, and the polar angle at that point, $\mathnormal{\phi}$.  Since the X-plane is defined 
by a single track, a unique solution in this space is extracted.

\begin{figure}
\begin{center}
\scalebox{1.0}{\includegraphics{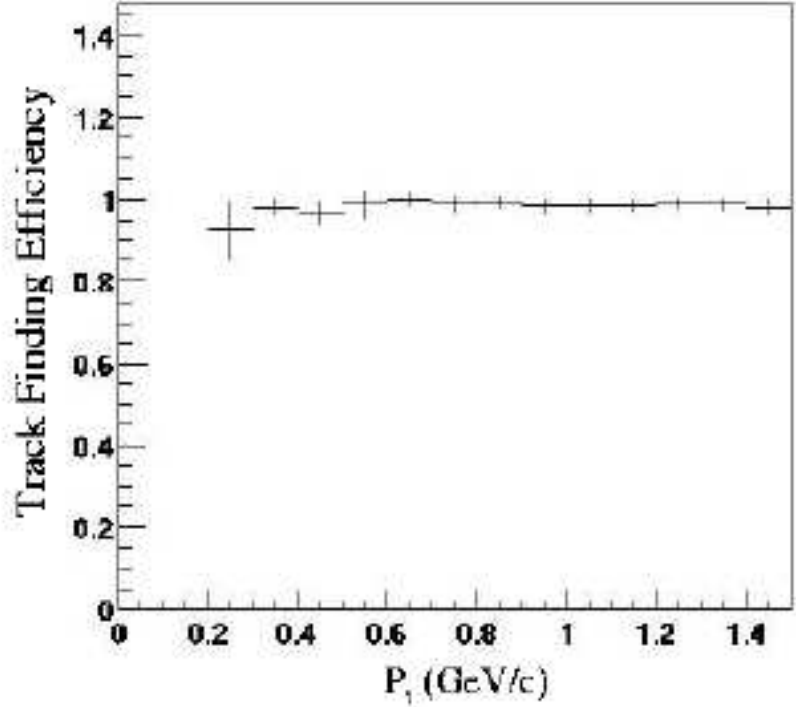}}
\end{center}
\caption{The reconstruction efficiency in HIJING-simulated central Au+Au events as a function
of the transverse momentum of the track.}
\label{dceff}
\end{figure}

The method of determining the drift chamber track reconstruction efficiency will be discussed 
in more detail in Section 12.  Using this method, the drift chamber tracking efficiency as 
a function of transverse momentum is shown in Fig. ~\ref{dceff} with less than 1\% spurious (ghost) tracks.
The resulting spatial resolution of the tracks is comparable to the single-hit resolution of 
the detector.

\section{Time Expansion Chamber}

Track reconstruction in the Time Expansion Chamber (TEC) is performed using a Combinatorial
Hough Transform method using the identical variables, $\mathnormal{\alpha}$ and 
$\mathnormal{\phi}$, and with an identical method for track reconstruction in the magnetic 
field bend plane as the drift chamber track reconstruction.

\begin{figure}[htb]
\begin{minipage}[t]{140mm}
\scalebox{0.7}{\includegraphics{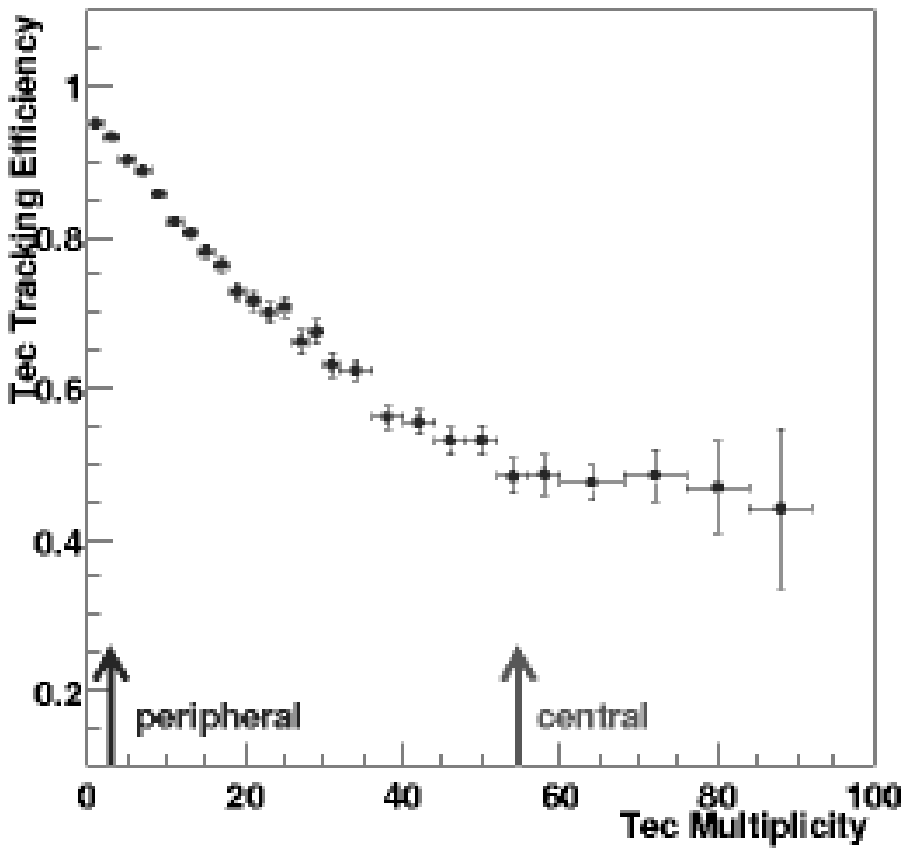}}
\hspace{\fill}
\scalebox{0.7}{\includegraphics{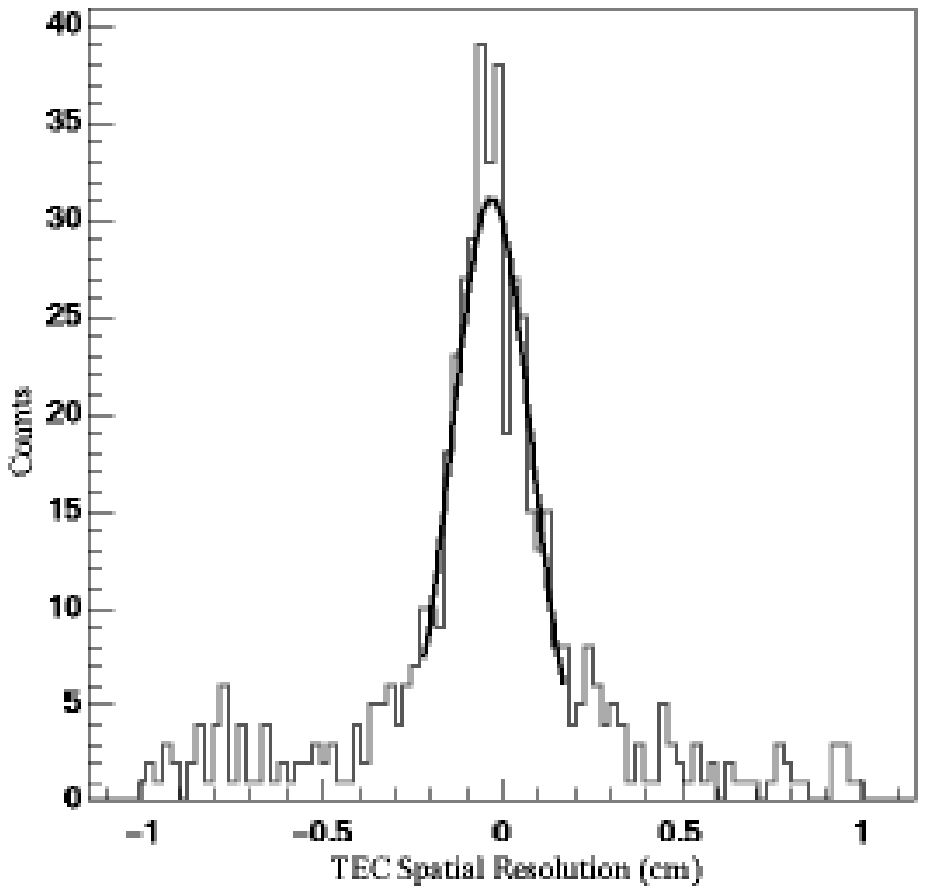}}
\vspace{0.1cm}
\caption{The TEC track reconstruction efficiency as a function of the TEC charged track
multiplicity obtained by evaluating simulated tracks embedded in real data events (left
panel), and the spatial resolution of 380 $\mu$m of the TEC based upon data taken in Run 2000
(right panel). The arrows on the left panel correspond to the 5\% most central, and the 60-80\%
peripheral events in the Au+Au collisions.}
\vspace{0.5cm}
\label{tecperform}
\end{minipage}
\end{figure}

A detailed simulation of the TEC detector response is used to determine the TEC tracking
efficiency.  This simulation calculates the average number of primary ion collisions for each
track section that traverses a single drift cell \cite{erm69}.  Fluctuations of the number of
primary collisions are modelled by Poisson statistics \cite{sau77}.  Secondary ionization is
calculated by generating a spectrum of $\mathnormal{\delta}$ electrons \cite{sau77}.  The total
number of primary and secondary ion pairs are calculated and the distance of the electron
cluster to the wire, along with the drift time, are calculated assuming a straight line 
drift.  Gas gain fluctuations are simulated using a {\em polya} distribution \cite{bel94}.
After the addition of noise, and the addition of the electronics shaping time response, 
the charge collected on the wire as a function of time bin is calculated.  The charge 
is finally converted into FADC channels for each track.  The final FADC contents are obtained by 
summing the contributions from all charged tracks in the event. The track reconstruction 
efficiency is determined by embedding individual simulated tracks processed through the 
response simulation into a real data event, reconstructing the resulting
{\em merged} event and determining how many of the embedded tracks are correctly
reconstructed.  The reconstruction efficiency is shown as a function of the charged
track multiplicity in the TEC in the left panel of Fig. ~\ref{tecperform}.  The spatial 
resolution of reconstructed tracks in the TEC, shown in the right panel of Fig. ~\ref{tecperform}, 
is determined directly from data taken in Run 2000 by comparing the residuals of tracks reconstructed 
independently in alternating planes of the TEC and fitting them to a Gaussian curve.  The resulting 
resolution, also confirmed within the simulation procedure, is determined to be 380 $\mu$m.

\section{Electromagnetic Calorimeters}

\begin{figure}
\begin{center}
\includegraphics*[width=12cm]{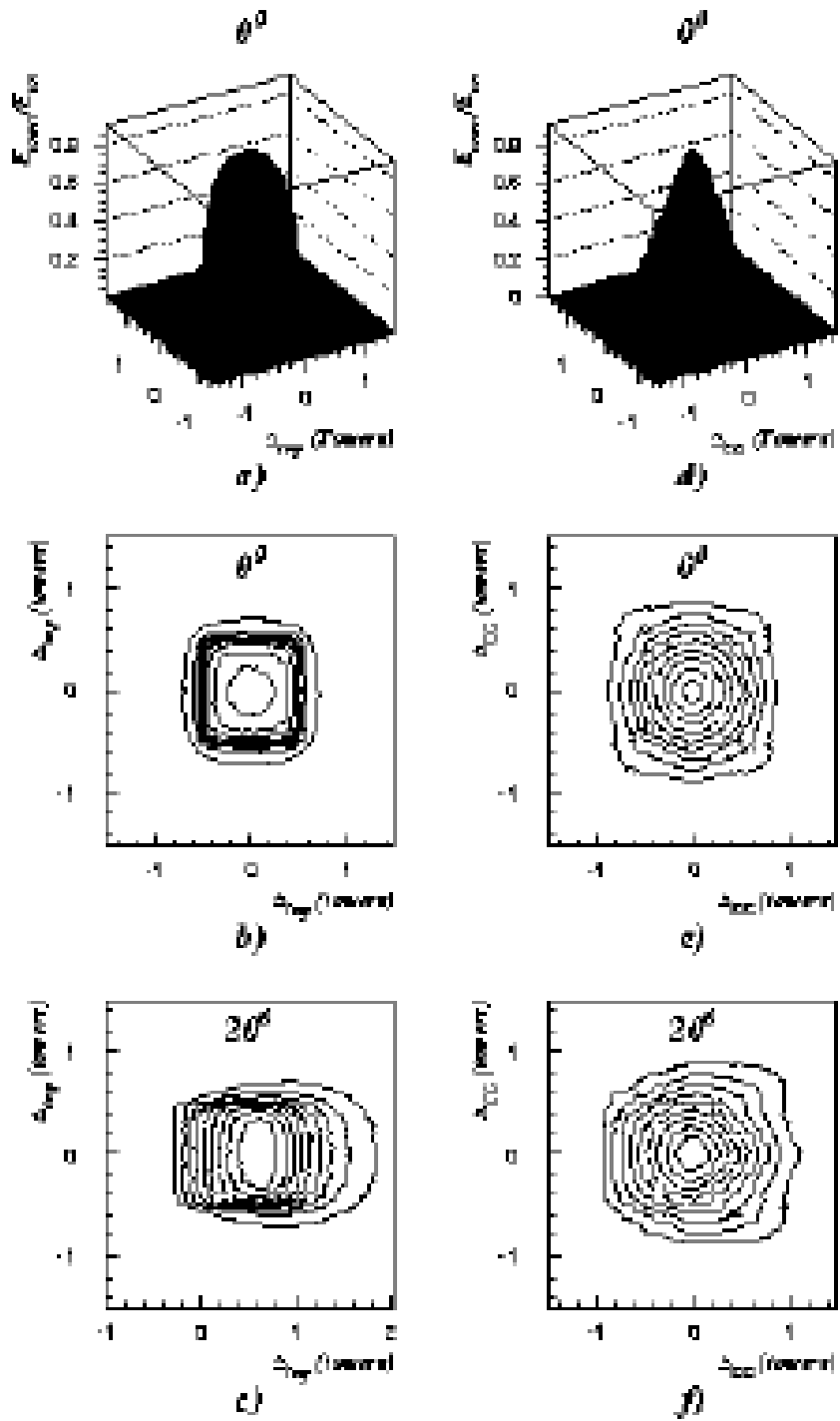}
\end{center}
\caption{Energy deposited in a PbSc calorimeter tower as a function of the distance 
between the tower center and the impact point (left panels), the tower center and  the
shower center-of-gravity (right panels); a),b),d) and e) are for orthogonal 
impact; c) and f) are for a $20^0$ impact angle.}
\label{emcshower}
\end{figure}

Pattern recognition in both the PbSc and PbGl Electromagnetic Calorimeters (EMC) includes single 
electromagnetic shower identification including calculation of the shower position, energy,
time-of-flight, overlapping shower separation, and hadronic shower rejection. The technique is 
based upon matching energy- and impact-angle-dependent descriptions of the projected electromagnetic 
shower shape as measured in a test beam of electrons in the momentum range 
0.3-5.0 GeV/c \cite{dav96} and reproduced in a detailed GEANT simulation of the detector response.  
An illustration of the shape of a shower is shown in Fig. ~\ref{emcshower} for a particle impact orthogonal 
to, and at a $20^0$ impact angle to, the calorimeter surface.

The shower reconstruction algorithm includes two steps. First, a cluster search is performed. The cluster 
is defined as a contiguous group of calorimeter towers with measured energies above a certain threshold. 
At this point, the clusters are considered to be totally independent. Second, the clusters are split 
into subclusters associated with local maxima. A local maximum is defined as a tower which has a 
deposited energy in excess of the energies measured in any of the 8 surrounding towers. A subcluster 
(peak region) is formed from the towers of the cluster sharing the energy in each tower among photons located 
in each peak region. Here, an iterative procedure based on a parameterized shower profile 
is used \cite{led93}.  Each subcluster is assigned a $\chi^2$ value for shower identification purposes.
 
The simplest algorithm for determining the particle position in a cellular calorimeter is to measure 
the center-of-gravity of the shower, $X_{CG}$: 
\begin{equation}
X_{CG} \equiv \frac{\sum{x_i E_i}}{\sum{E_i}}, \label{eq:emc.pos.cg}
\end{equation}
where $x_i$ is the $x$-coordinate of the center of tower $i$, and $E_i$ is the energy deposited 
in the tower.  For electron and photon position reconstruction, a parametrization of the shower 
center-of-gravity in units of tower width is applied \cite{baz98}: 
\begin{equation}
X_{CG} = \frac{1}{2}\frac{sinh(\frac{X_{imp}+\Delta-\delta}{b})}{sinh(1/2b)}+\delta; 
|X_{imp}+\Delta-\delta|\leq\frac{1}{2}. \label{eq:emc.pos.cg1}
\end{equation}
where $X_{imp}$ is the particle impact point, $b$ is the shower cross-sectional width, 
$\Delta$ is the mean displacement of the calculated shower center-of-gravity
from the impact point $X_{imp}$, and $\delta$ is a phase shift related to the skewed 
shape of the shower projection. All of these parameters ( $b$, $\Delta$, and $\delta$) 
are parameterized as a function of the incident particle energy, $E$, and angle of incidence,
$\theta$. $\Delta$ is given by
\begin{displaymath}
\Delta = L_{eff} \cdot sin(\theta), \label{eq:emc.delta}
\end{displaymath}
where $L_{eff}$ is the effective depth of the shower penetration in the calorimeter as
determined by the position of the cascade-curve median in the longitudinal direction 
($\propto ln(E)$) \cite{gro00}. The value of $b$ is given by
\begin{displaymath}
b = b_0 + b_1(E)sin^2(\theta), \label{eq:emc.b}
\end{displaymath}
where $b_0$ is a constant and $b_1(E)\propto ln(E)$.
$\delta$ is found to be essentially energy-independent and can be parameterized as a 
function of the impact angle only. Finally, the position of the particle impact on the calorimeter 
surface, $X_{imp}$, is calculated according to the formula obtained by inverting Equation
(\ref{eq:emc.pos.cg1}): 
\begin{displaymath}
X_{imp} = b \cdot arcsinh[2 \cdot (X_{CG}-\delta) \cdot sinh(1/2b)]-\Delta+\delta, \label{eq:emc.ximp}
\end{displaymath}
\begin{displaymath}
|X_{CG}-\delta| \leq \frac{1}{2}.
\end{displaymath}
The calorimeter position resolution is well described by \cite{dav98,baz98}:
\begin{displaymath}
\sigma_X(E,\theta) = \sigma_X(E,0^0) \oplus (d \cdot sin(\theta)), \label{eq:emc.sigmax}
\end{displaymath}
where $\sigma_X(E,0^0)$ is the position resolution for an orthogonal 
incidence. The second term in this parametrization describes the 
contribution of the longitudinal shower fluctuations on the projection to
the calorimeter surface. The value of $d$ is
$0.8X_0$, where $X_0$ is the radiation length in the calorimeter module. 
Using the electron test beam, a position resolution of $5.7 mm/\sqrt{E(GeV)} + 1.6 mm$ was achieved.

A number of factors are taken into account when reconstructing the photon and electron
energy in the calorimeters: the longitudinal energy leakage, the light attenuation in 
the calorimeter module, the response dependence on the particle impact position, 
and the shower overlaps in the high multiplicity environment, some of which will be
discussed here. The nonlinearity of photon and electron energy measurements in the 
calorimeters is primarily connected with light attenuation in a calorimeter module.
This is well described by the expression $E_{EMC}/E \approx E^{\frac{X_0}{\lambda_{att}}}$, 
where $E_{EMC}$ is the energy measured in the calorimeter, and $E$ is the actual photon or 
electron energy in GeV.

In a high multiplicity environment, shower overlaps can seriously affect the electromagnetic shower 
energy measurements, thus degrading the energy resolution and inducing systematic shifts in the 
measurement.  For a tower energy threshold of 3 MeV, the mean cluster size in the PbSc calorimeter 
initiated by a 0.5 GeV (4 GeV) photon is 7 (19) towers. The larger the cluster size, the higher 
the probability for an overlap. However, the shower size can be reduced to 4-5 towers in the 
shower core. Towers belonging to the shower core are defined using a parameterized shower profile: 
\begin{displaymath}
r = \frac{E_i^{pred}}{E} < \epsilon, \label{eq:emc.r}
\end{displaymath}
where $r$ is the contribution of the tower to the total shower energy $E$, and $E_i^{pred}$ is the 
predicted energy deposited in the tower. The $\epsilon=0.02$ parameter is chosen in order to have 
as few towers in a cluster as possible while keeping the energy resolution on the same level.  
The shower core defined in this manner contains about $92\%$ of the total shower energy, 
dropping to $90\%$ for a $20^0$ impact angle. Using the shower core for energy measurements 
introduces an additional constant term of about $2\%$ in the energy resolution, which slightly worsens 
the energy measurements of isolated photons and electrons, but helps to considerably improve the 
energy measurements in the high multiplicity environment, where an energy resolution for the
PbSc calorimeter of $8.2\%/\sqrt{E(GeV)}\oplus1.9\%$ was achieved in electron test beams at the BNL
AGS and the CERN SPS \cite{dav96}.

The time-of-flight (TOF) information provided by the calorimeters (using the BBC as the start counter)
is a useful tool for particle identification.  This helps by enhancing the $e/\pi$ separation at 
low momenta ($<0.8$ GeV/c), where the energy-to-momentum comparison, as well as the shower 
profile analysis, lose effectiveness for $e/\pi$ discrimination.  In addition, this is the primary 
tool for identifying neutral hadrons, which are a major source of background in the photon spectrum 
in the energy region of approximately 1 to 3 GeV.  Also, the magnet poles and structural 
elements of the detectors in front of the calorimeter are sources of non-vertex photons, part of which 
($\sim15$\%) can be rejected using the TOF information.  Finally, in the case of high multiplicity 
events, requiring consistency of the TOF values measured in the different towers within the same cluster 
can help to identify accidental shower overlaps. Several corrections are applied to measure the TOF, 
including slewing corrections and flight path corrections based upon the collision vertex, after 
which a resolution of better than 700 psec for the PbSc calorimeter was achieved.

\section{Ring Imaging Cherenkov Detector}

\begin{figure}
\begin{center}
\includegraphics*[width=10cm]{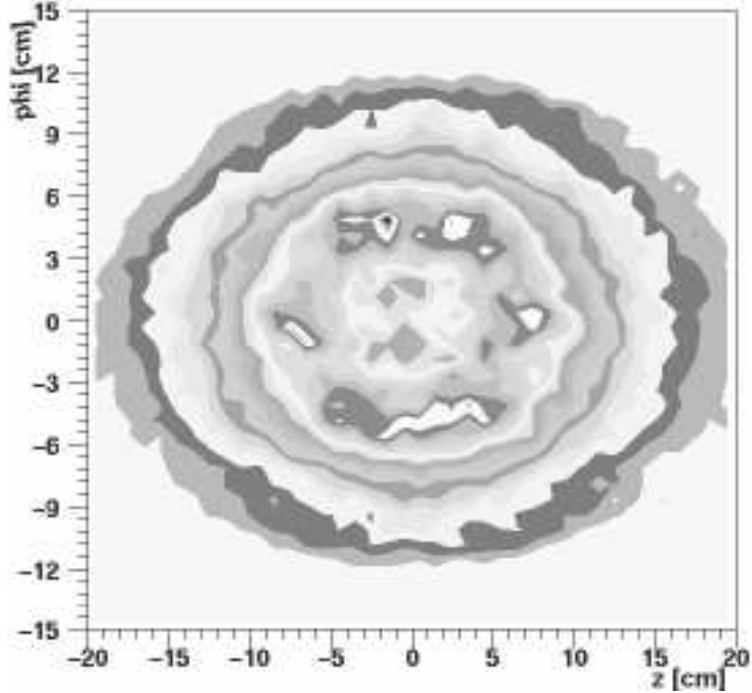}
\end{center}
\caption{The distribution of hit PMT's around the projected ray from a reconstructed track
in zero-field runs after the geometry calibration.}
\label{crkring}
\end{figure}

Electron identification using the Ring Imaging Cherenkov Detector (RICH) is performed with two
C++ classes. The first class handles the geometry of the RICH detector by modelling the 
position, shape, and orientation of the 96 mirror panels and the position of the 5120 PMT's 
in the RICH detector. This class also handles all optical tracing calculations. The second
class uses the geometry class and performs the electron identification. The electron 
identification class accepts a previously reconstructed track and searches for RICH PMT's that can be 
associated with the track. This is done by first reflecting the track about the RICH mirror 
as if it were a light ray, and then projecting the reflected track onto the RICH PMT 
array.  If the track is an electron, the reflected ray should be projected to the center 
of a ring formed by hit PMT's. The number of hit PMT's within a tight association radius from 
the projected track is counted and the number of photo-electrons measured in the PMT's in 
the association radius is summed. In addition, a quality factor based upon the position 
of the associated hits and the position of the projected track is calculated. All of these
quantities are then used to separate electron tracks from the hadron background.

\begin{figure}
\begin{center}
\includegraphics*[width=11cm]{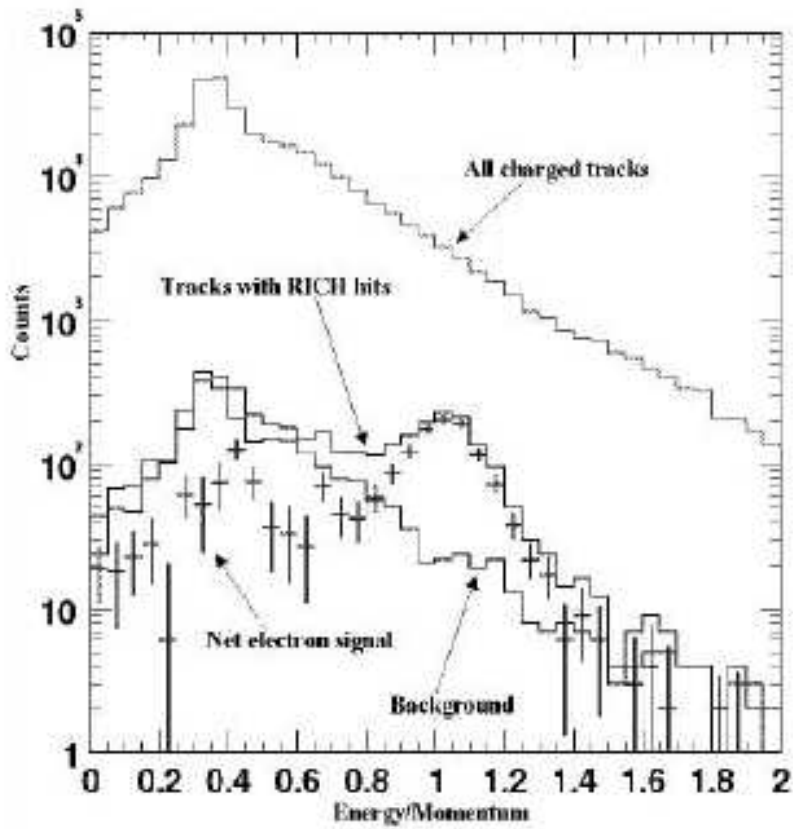}
\end{center}
\caption{
The electron signal observed in PHENIX. The highest histogram is the distribution of 
the energy and momentum ratio of all charged tracks in $0.8<p_t<0.9$ GeV/c. The middle 
histogram is the one with the RICH hits. The lowest histogram is the distribution of a random
 background estimated from the data. The crosses are the difference of the middle and the
lowest, resulting in the net electron signal. The electron signal below $E/p<0.8$ are 
caused by electrons from photon conversions far from the beam pipe.}
\label{crkelectron}
\end{figure}

The gain of the RICH PMT's is determined from the pulse height of the single photo-electron 
peak in the data. The geometry of the mirror system and the optics are calibrated from the data 
using zero magnetic field runs. In these runs, charged particle tracks are reconstructed as a 
straight line connecting the event vertex (determined by the BBC), the PC1 hit position, 
and the PC3 (or EMCAL) hit positions. The orientation of the mirror panels in the geometry 
class are then adjusted so that the projection point of the track reflected by the mirrors 
points to the center of the Cherenkov rings recorded in the PMT array. Fig. ~\ref{crkring} shows the 
distribution of the hit PMT's around the projected ray after the geometry calibration.

The hadron rejection factor provided by the RICH strongly depends upon the multiplicity of 
the event. From data taken with a test beam with a small proto-type of the RICH detector, 
it was shown that the RICH detector can provide a pion rejection factor better than 
$10^4$ with an electron efficiency close to 100\%. However, in high multiplicity events, the
electron-pion separation is reduced since a hadron track can be associated with a RICH hit 
produced by background electron tracks, resulting in a track that is mis-identified as an
electron. The pion rejection factor in most central Au+Au collisions was determined to be
on the order of several hundred with an electron efficiency of $\sim80\%$. Fig. ~\ref{crkelectron}
illustrates electron identification using the RICH detector. In this figure, the ratio of the energy 
(measured by the calorimeters) and the momentum of the charged tracks in the $p_t$ range
$0.8 < p_t < 0.9$ GeV/c are plotted. For all charged tracks, the $E/p$ ratio has a {\em mip}
peak at $E/p\sim0.3$ on a broad distribution caused by hadronic interactions. When RICH hits
are required, a clear peak appears at $E/p=1.0$, which is the electron signal. The 
background due to random associations of the RICH and the track is estimated, and also shown 
in Fig. ~\ref{crkelectron}.

\section{Time-Of-Flight Detector}

\begin{figure}
\begin{center}
\includegraphics*[width=14cm]{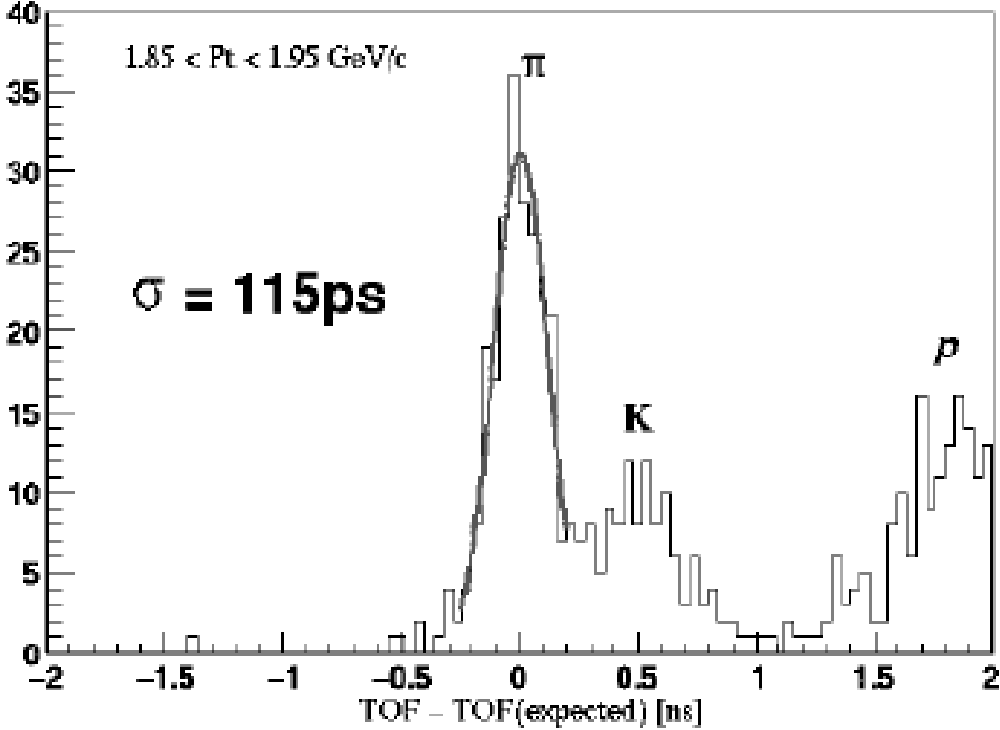}
\end{center}
\caption{The time-of-flight resolution in the transverse momentum range $1.85 < p_t < 1.95$ GeV/c 
for positively charged pions. An overall time-of-flight resolution of 115 ps is achieved.}
\label{tofres}
\end{figure}

Prior to the application of the Time-Of-Flight detector (TOF) for  particle identification,
the flight time of a particle impacting the TOF is determined by taking 
the average of the measured stop time values at both ends of the hit PMT after subtraction of 
the start timing provided by the BBC. The pulse height information from the PMT's is used to
reconstruct the hit position by calculating the center-of-gravity of the pulse heights from each
end of the slat after applying timing calibrations, gain calibrations, geometrical alignment 
corrections, and pulse-height-dependent timing, or {\em slewing effect}, corrections.

For TOF event reconstruction, four distinct C++ classes are used. The first class defines an 
address object, which provides the mapping between the actual read-out channel number of the
Front End electronics Module (FEM) and a sequential slat identification number. The second 
class defines a geometry object, which handles the conversion from the slat identification number
to the geometrical hit position of the particle within the PHENIX coordinate system. The third class 
defines a calibration object, which deals with all of the calibration parameters used for event 
reconstruction for the TOF system, such as the slat-by-slat timing offset, the position offset,
and the PMT gain. Finally, the address, geometry, and calibration objects are accessed within a 
TOF reconstruction object within which the energy loss in the scintillators and the hit positions 
are reconstructed.  

During Run 2000, the TOF detector timing offset was calibrated on a slat-by-slat basis using 
reconstructed track information. An overall time-of-flight resolution of $\sim$110-120 ps is achieved 
by selecting pions at high momentum and comparing their time-of-flight to that expected from a $v=c$
particle with a linear trajectory originating at the event vertex, as shown in Fig. ~\ref{tofres}
after corrections for the slewing effect and timing drift during the run.  The method of particle 
identification using the TOF detector will be discussed in Section 13.

\section{Momentum Reconstruction and the Track Model}

Due to the complicated, non-uniform shape of the focusing magnetic field along the flight 
path of charged particles traversing the PHENIX central arm spectrometers, an analytic 
solution for the momentum of the particles cannot be determined. Therefore, other approaches
such as look-up tables, must be used \cite{chi96}.  For Run 2000, a four-dimensional 
field-integral grid was constructed for momentum reconstruction using the drift
chamber and for track model definition within the entire radial extent of the central arms.  The 
variables in the field-integral grid are the $z$ coordinate of the event vertex; the polar angle, 
$\theta_0$, of the particle at the vertex; the total momentum of the particle, $p$; and the 
radius, $r$, at which the field-integral $f(p,r,\theta_0,z)$ is calculated.  The field-integral 
grid is generated by explicitly swimming particles through the measured magnetic field map and 
numerically integrating to obtain $f(p,r,\theta_0,z)$ for each grid point.

An iterative procedure is used to reconstruct the momentum of a reconstructed track, utilizing
the fact that $f(p,r,\theta_0,z)$ varies linearly with the $\phi$ angle of the track at a 
given radius. This can be expressed as $\phi = \phi_{0} + q \cdot f(p,r,\theta_{0},z)/p$.
Each track is assumed to be a primary track originating from the event vertex as determined 
by the BBC. An initial estimate of the track momentum and charge is made from the reconstructed 
bend angle, $\alpha$, of the track in the drift chamber.  The measured polar angle, $\theta$, 
of the track in the non-bend plane at the drift chamber reference radius is used as an initial 
estimate of $\theta_0$.  Then, using the radial position of each reconstructed hit associated
to the track, a four-dimensional polynomial interpolation of the field-integral grid is performed 
to extract a value of $f(p,r,\theta_0,z)$ for the drift chamber hit.  Once this is done for all 
hits, a robust fit in $\phi$ and $f(p,r,\theta_0,z)$ is performed to extract the quantities 
$\phi_0$ and $q$/$p$ for the track. The extracted values are then fed back into the above equation.
The initial polar angle, $\theta_{0}$, is also determined using an iterative procedure using the 
equation $\theta = \theta_{0} + \delta (p,r = R, \theta_{0},z) - g(p,r = R, \theta_{0},z)/p$
where $\delta$ is the bend angle of the particle trajectory relative to the straight-line 
trajectory of an infinite-momentum particle.  Typically, less than four iterations are necessary 
for convergence on these quantities.

\begin{figure}
\begin{center}
\scalebox{0.65}{\includegraphics*[width=14cm]{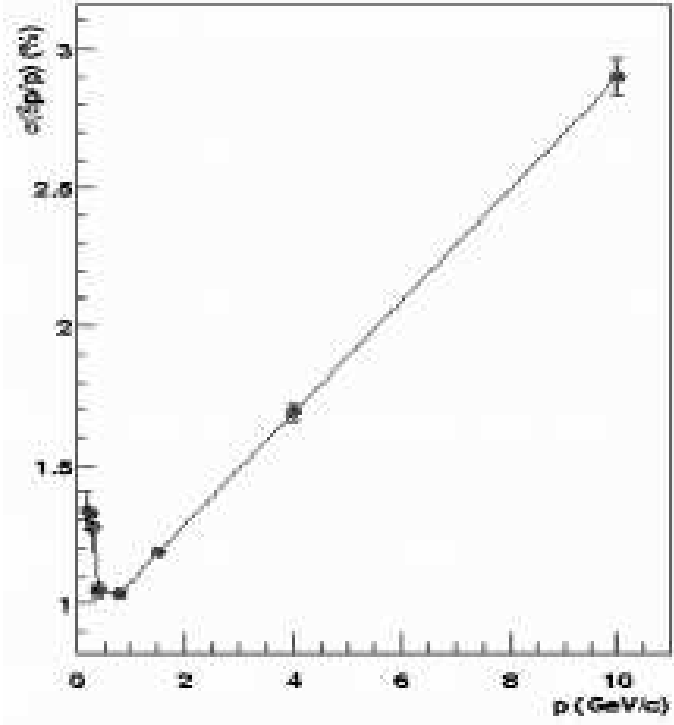}}
\end{center}
\caption{Momentum resolution as a function of momentum from simulated single particle events.}
\label{momres}
\end{figure}

This procedure has an additional advantage in that it can be used to define the shape of the
track within the central arm magnetic field, which can be then be used to determine the track 
intersections with each detector in order to facilitate inter-detector hit association as 
described in the next section. This is done by storing the coordinates of the particle in radial 
steps as additional entries (but not keys) in the field-integral grid.  Line segments connecting
the interpolated grid coordinates for a track are intersected with the geometry objects describing 
the position of each detector in order to estimate the projection of the track on each detector.
These projections are used as a seed for inter-detector hit association.  Finally, the length of
the interpolated line segments from the event vertex to a given detector can be summed in order
to provide an estimate of the flight distance of the particle to that detector.  This quantity is
used to facilitate particle identification using the TOF.

The momentum reconstruction resolution is determined by generating simulated muons (which will 
not decay as they traverse the spectrometer) with momentum $p_{gen}$ and event vertices ranging 
from $-40 < z < 40$ cm. A complete GEANT and detailed detector response simulation on each track 
is performed to obtain the detector hits as they would appear in the data, after which the
reconstruction is run in the same manner as for data, producing a reconstructed momentum, $p_{rec}$.
The quantity $|p_{rec}-p_{gen}|/p_{gen}$ (in percent) is shown in Fig. ~\ref{momres}. Multiple 
scattering dominates the resolution at low momenta, while detector resolution determines the 
resolution at higher momenta. In addition, the momentum resolution is not multiplicity-dependent, 
as observed by generating the same plot using central Au+Au HIJING events.

\section{Inter-Detector Association}

\begin{figure}
\begin{center}
\rotatebox{-90.0}{\scalebox{1.0}{\includegraphics{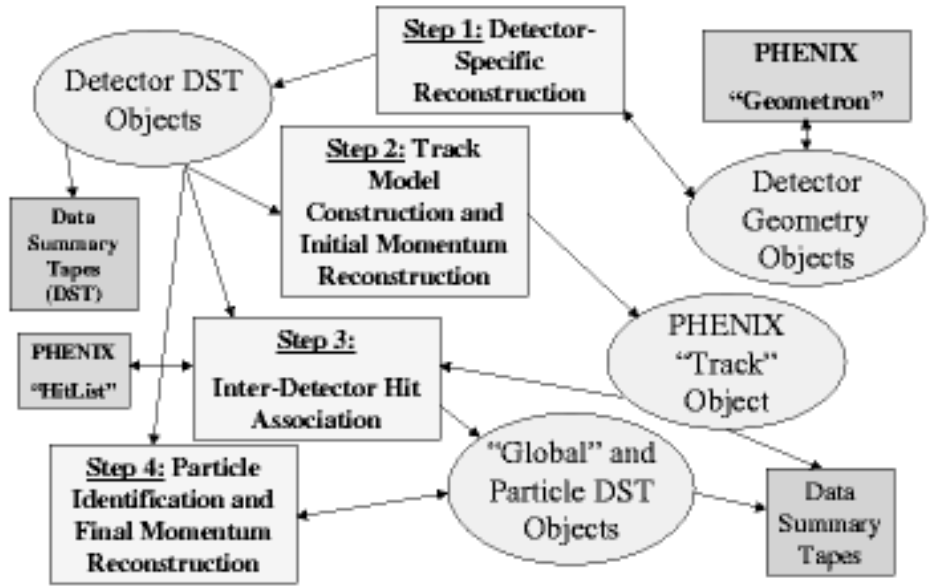}}}
\end{center}
\caption{Illustration of the object-oriented approach to inter-detector association in PHENIX.}
\label{ooflow}
\end{figure}

This section will address the methods and algorithms used for inter-detector association.
The general object-oriented flowchart used for inter-detector, or {\em global}, reconstruction
is shown in Fig. ~\ref{ooflow}.  All stages of the event reconstruction query the geometry information for
the spectrometer, which is stored as {\em Detector Geometry Objects}.  Geometry interactions 
(intersections, transformations, etc.) are handled either within a specific geometry object type, 
or within a C++ singleton class called the {\em Geometron}.  The output of detector-specific 
reconstruction are stored in objects containing the hit and track information, which are stored 
in the Data Summary Tapes (DST) created after each reconstruction pass.  The anchor of inter-detector
hit association is the track model object, which is discussed further below.  Once a track model 
object is created, its methods are used to perform hit association in addition to momentum reconstruction 
with the assumption that the particle is a pion.  The resulting association and kinematics 
information is stored in distinct objects (and in the DSTs), which are then fed into the particle 
identification procedures.  For particles that have a probability to not be a pion, the momentum 
can be subsequently reconstructed using the updated particle assumption, as its momentum will differ
from that of the pion due to energy loss as it traverses the spectrometer.

The track model object is implemented in a flexible manner through a base class named {\em PHTrack}, 
which defines the basic functionality that must be provided by a generic PHENIX track model class.  
The momentum reconstruction algorithm previously described is implemented as a class that inherits
from the {\em PHTrack} class.  The most important application of the track model is to determine 
its intersection with the various detectors in order to perform inter-detector track and hit 
associations.  Thus, the data members of the {\em PHTrack} base class include a list of projection 
points, vectors, and errors at each detector.  Also included is a list of three-dimensional coordinates 
approximating the shape of the track for use in event displays.  It is up to the individual track 
models that inherit from {\em PHTrack} to provide the methods for the construction and projections of 
a track prior to associating hits to the track.  The basis for the construction of a track model
object is typically, but not restricted to, a drift chamber track. During Run 2000, two distinct 
track models were utilized.  The first was an analytical linear track model used for data taken 
with the magnetic field turned off, and the second was the one used for momentum reconstruction.

The inter-detector hit association algorithm operates on each {\em PHTrack} object within an
event individually.  Additional inputs for hit association include lists of the detector hits 
(or tracks) which are to be associated to the {\em PHTrack} object.  These ``hit lists'' are 
constructed and sorted by spectrometer arm and by increasing azimuthal angle using a class developed 
for this functionality in order to simplify and speed up processing.  Once constructed, the manipulation 
of these ``hit list'' objects are independent of the detector to which the hits or tracks belong.
Due to the small number yet wide resolution disparity of detectors that must be associated, 
the hit association is performed using a three-dimensional road algorithm with each detector 
associated independently.  For a given {\em PHTrack} object, its projection point at the detector 
being associated is used to define the center of a window spanning the $\phi$  and $z$ coordinate 
directions.  Projection error estimates from the {\em PHTrack} object, which can be momentum-dependent,
are then used to define the widths of the roads in $\phi$ and $z$.  These widths can be scaled at 
run-time via user input parameters, typically after examining the residuals of the projections.
Once the window is defined, it's ranges are fed to the ``hit list'' object, which returns either 
a list of all hits within the window, or the closest hit to the center of the window within its 
boundaries.  There is one special case of TEC track association considered in this algorithm.  
For this detector, the $\alpha$ angle of the TEC track, along with the $\alpha$ angle of the drift 
chamber track, are correlated with the inverse $p_t$ of the track.  This information can be used to apply 
an $\alpha$-difference angle cut between the two detector tracks to reduce the list of TEC tracks 
being considered for association.  For the analysis of the data taken in Run 2000, the accuracy of 
the projection determination was good enough that the closest hit was used in the initial 
reconstruction pass while retaining the option for specific analyses to further optimize the 
association hit set in later passes on the DST's using the ``hit list'' methods.

\begin{figure}[htb]
\begin{minipage}[t]{140mm}
\scalebox{1.0}{\includegraphics{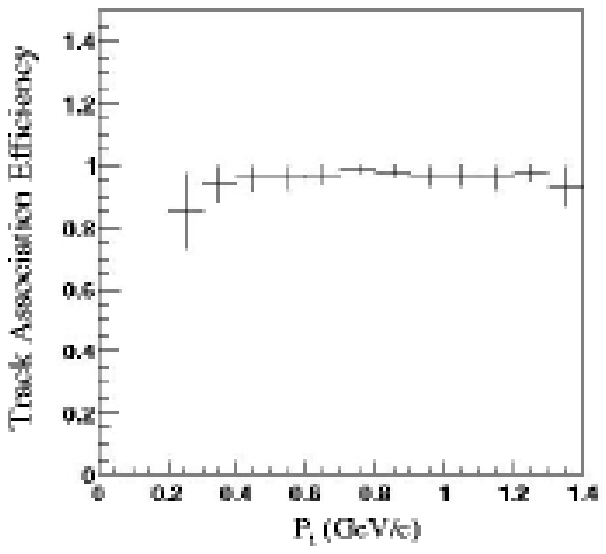}}
\hspace{\fill}
\scalebox{1.0}{\includegraphics{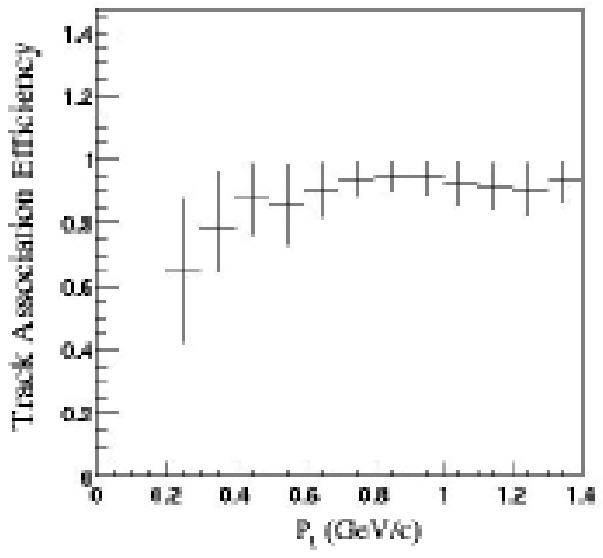}}
\vspace{0.1cm}
\caption{The hit association efficiency between the drift chamber and PC1 (left), and the drift chamber
and PC3 (right) as a function of $p_t$ in central HIJING 200 GeV/A Au+Au simulated events.}
\vspace{0.5cm}
\label{gloeff}
\end{minipage}
\end{figure}

The efficiency of the inter-detector hit association, with the association being defined as the 
hit closest to the center of the projection window, has been calculated based upon 
200 GeV/A Au+Au central events generated using the HIJING event generator \cite{wan91}.
These events were processed through the GEANT-based simulation of the detector response for 
all detectors in the central arm spectrometers.  The output of the response simulation was then 
reconstructed in the same manner as the data while storing the pointers that relate reconstructed 
global tracks back to the GEANT tracks that contribute to them.  Utilities that trace a 
reconstructed object back to the GEANT information are implemented as a singleton class for 
use by a wide variety of evaluations.  For the evaluation of the global association, the 
{\em dominant contributor} of the drift chamber track was used as a basis.  The {\em dominant contributor}
is defined as the input GEANT track which contributes the most hits to the reconstructed track.  
In order for an association to be considered to be made correctly, the GEANT track that provided the
reconstructed hit or track for the detector in question must match the dominant contributor of 
the reconstructed drift chamber track to which it is associated.  The efficiency is defined as the ratio of 
the number of tracks that were correctly reconstructed and geometrically reconstructable to the number of 
tracks that were geometrically reconstructable by each detector (in order to isolate the efficiency to that 
of the association only).  The results of the efficiency calculation are shown in Fig. ~\ref{gloeff} for 
associations to PC1, which is situated adjacent to the drift chamber, and for associations to PC3 after spanning the 
tracking-detector-free gap occupied by the RICH.  Although the efficiencies in the left panel of
Fig. ~\ref{gloeff} are lower, the performance is such that good particle identification can be 
achieved using the TOF, as demonstrated in the next section.

\section{Particle Identification Using the Time-Of-Flight Detector}

\begin{figure}
\begin{center}
\includegraphics*[width=11cm]{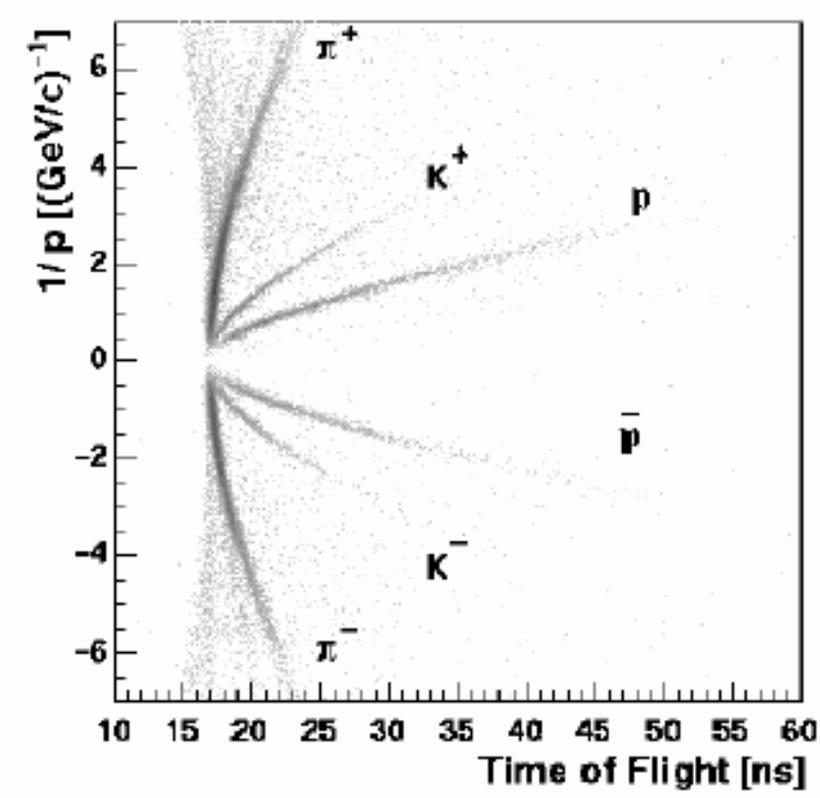}
\end{center}
\caption{
Scatter plot of the time-of-flight versus reciprocal of the momentum in minimum bias 
Au + Au collisions. This demonstrates the clear particle identification capability 
using the TOF in the Year 2000 data taking period. The flight path is corrected 
assuming the mass for each particle species. 
}
\label{tofpid}
\end{figure}

Particle identification for charged hadrons is performed by combining the 
information from the drift chamber, PC1, BBC, and the TOF.  The time-of-flight 
resolution of the TOF is measured to be 115 ps, resulting in PID capability for high 
momentum particles. A 4$\sigma$ $\pi$/K separation at momenta up to 2.4 GeV/c, and a 
K/proton separation up to 4.0 GeV/c can be achieved as shown in Fig. ~\ref{tofpid}.

\begin{figure}
\begin{center}
\scalebox{1.0}{\includegraphics{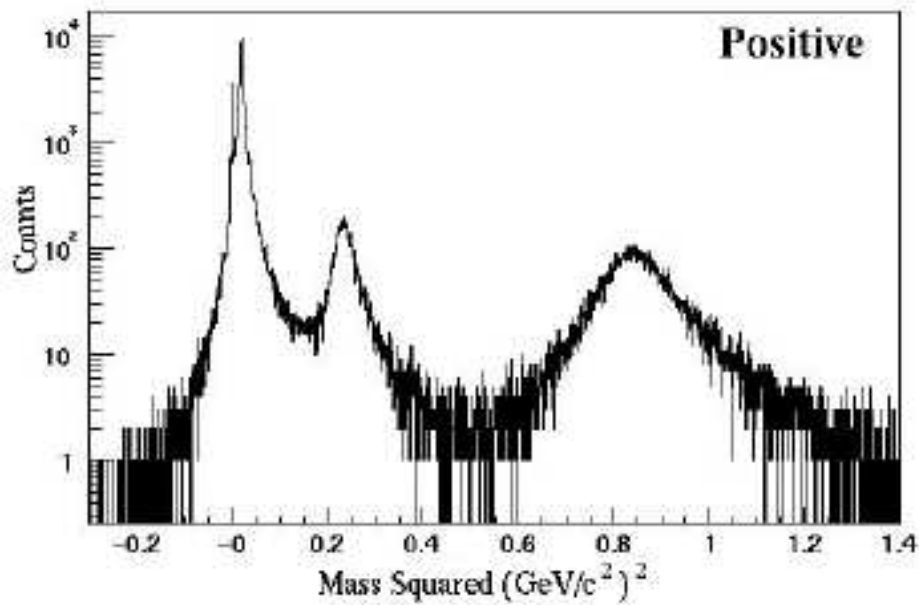}}
\end{center}
\caption{
The mass squared distribution for positively charged particles integrated over
all momenta.
}
\label{tofmass2}
\end{figure}

Once a global track is reconstructed with information from the drift chamber, PC1, 
and TOF (using a window for TOF association adjusted so that the residuals between the 
projection point and the reconstructed TOF hit position is within a standard deviation 
of 2.5), the flight-path length of the track from the event vertex to the TOF as calculated 
by the momentum reconstruction algorithm is used to correct the time-of-flight value 
measured by the TOF.  Fig. ~\ref{tofpid} shows a scatter plot of corrected time-of-flight as a function 
of the reciprocal of the momentum in minimum-bias Au+Au collisions, after the momentum-dependent 
residual cut between the track projection point and TOF hits. The flight path is also 
corrected for each particle species in this figure. Fig. ~\ref{tofmass2} shows the mass-squared 
distribution for positively charged particles integrated over all momenta. The vertical axis in
this figure is in arbitrary units. It is demonstrated that clear 
particle identification using the TOF is achieved in Run 2000. 

\section{Summary}

This document has described the methods and algorithms used for event reconstruction in the
PHENIX central arm spectrometers.  Using a modular, object-oriented implementation, widely
disparate algorithms from widely disparate detector types have been integrated into a coherent 
reconstruction chain in order to successfully reconstruct hits and tracks within each detector, 
associate hits from different detectors, reconstruct particle momenta, and provide particle
identification for PHENIX physics analysis.

\section{Acknowledgements}

We thank the staff of the RHIC project, Collider-Accelerator, and Physics
Departments at BNL and the staff of PHENIX participating institutions for
their vital contributions.  We acknowledge support from the Department of
Energy and NSF (U.S.A.), Monbu-sho and STA (Japan), RAS, RMAE, and RMS
(Russia), BMBF and DAAD (Germany), FRN, NFR, and the Wallenberg Foundation
(Sweden), MIST and NSERC (Canada), CNPq and FAPESP (Brazil), IN2P3/CNRS
(France), DAE (India), KRF and KOSEF (Korea), and the US-Israel Binational
Science Foundation.

\end{document}